\documentclass{aastex}

\usepackage{emulateapj5}

\newcommand{\et}{et al.}

\newcommand{\fv}{F_{var}}

\newcommand{\xte}{{\it RXTE}}

\newcommand{\xmm}{{\it XMM-Newton}}
\newcommand{\chandra}{{\it Chandra}}

\newcommand{\Msun}{\hbox{$\rm\thinspace M_{\odot}$}}
\slugcomment{}
\shorttitle{Markowitz}
\shortauthors{X-ray Variability of NGC 3783}

\begin{document}
\title{X-ray Variability Characteristics of the Seyfert 1 Galaxy NGC 3783}

\author{A.~Markowitz\altaffilmark{1,2}
\altaffiltext{1}{X-ray Astrophysics Laboratory, Code 662, NASA/Goddard Space Flight Center, Greenbelt, MD 20771; agm@milkyway.gsfc.nasa.gov}
\altaffiltext{2}{N.A.S./N.R.C.\ Research Associate}}

\begin{abstract}

We have characterized the energy-dependent X-ray variability properties of
the Seyfert~1 galaxy NGC~3783 using archival {\it XMM-Newton} and {\it Rossi 
X-ray Timing Explorer} data. The high-frequency fluctuation power spectral 
density function (PSD) slope is consistent with flattening towards higher 
energies. Light curve cross correlation functions yield no significant lags, 
but peak coefficients generally decrease as energy separation of the bands 
increases on both short and long timescales. We have measured the coherence 
between various X-ray bands over the temporal frequency range of $6 \times 
10^{-8} - 1 \times 10^{-4}$ Hz; this range includes the temporal frequency 
of the low-frequency power spectral density function (PSD) break tentatively 
detected by Markowitz \et\ and includes the lowest temporal frequency over 
which coherence has been measured in any AGN to date. Coherence is generally 
near unity at these temporal frequencies, though it decreases slightly as 
energy separation of the bands increases. Temporal frequency-dependent phase 
lags are detected on short time scales;  phase lags are consistent with 
increasing as energy separation increases or as temporal frequency decreases. 
All of these results are similar to those obtained previously for several 
Seyfert galaxies and stellar-mass black hole systems. Qualitatively, these 
results are consistent with the variability models of Kotov 
et al.\ and Lyubarskii, wherein the X-ray variability is due to inwardly propagating 
variations in the local mass accretion rate.
\end{abstract}
 

\section{Introduction}

Seyfert active galactic nuclei (AGNs) and stellar-mass black hole X-ray binary 
systems (XRBs) both exhibit rapid, aperiodic X-ray variability that likely 
originates in the innermost regions of these compact accreting objects. The 
dominant radiation at hard X-rays is generally thought to be inverse 
Comptonization of soft seed photons by a hot corona (e.g., Shapiro, Lightman 
\& Eardley 1976, Sunyaev \& Titarchuk 1980), though the exact geometry is 
uncertain and numerous models have been invoked (e.g., Zdziarski \et\ 2003). 
Possible configurations include, but are not limited to, a neutral accretion 
disk sandwiched by a patchy, and possibly outflowing, corona (e.g., Stern \et\ 
1995, Svensson 1996, Beloborodov 1999), as well as a hot inner disk radially 
surrounded by a cold disk, with a variable transition radius (e.g., Zdziarski, 
Lubinski \& Smith 1999).

The X-ray variability can be characterized by fluctuation power spectral 
density functions (PSDs) which show the "red-noise" nature of the variability 
at relatively high temporal frequencies in both AGNs and XRBs. In XRBs, the 
X-ray variability characteristics and PSD shape are known to correlate with energy 
spectral state. Modeling of the broadband PSDs of XRBs usually utilizes some combination 
of one, two, or more Lorenztians, which tend to dominate in the so-called low/hard energy
spectral state, plus a broadband noise component characterized as $P \propto f^{-1}$, 
which tends to dominates in the high/soft state (e.g., in Cyg X-1; see Axelsson \et\ 
2005, Belloni \et\ 2005). The peak frequency of the Lorentzians tend to move towards 
relatively higher temporal frequencies as luminosity and accretion rate increase in 
many transient sources (e.g., van der Klis 2004, Kalemci \et\ 2004, Rodriguez \et\ 
2004). The (relatively more stable) black hole candidate Cyg X-1 tends to spend a 
majority of its time in the low/hard state, wherein the PSD is dominated by two 
Lorentzians that typically peak near 0.1--0.3 and 3--10 Hz (e.g., Nowak \et\ 1999a,b). 
The PSD of the high/soft state of Cyg X-1 is dominated by 1/$f$ broadband noise at 
the lowest temporal frequencies, with a high-frequency break usually found near 10 Hz. 
Global accretion rate is the likely driver in determining whether an XRB appears in the
low/hard or high/soft state, with the transition usually occurring at $\sim2\%$ of Eddington
(Maccarone 2003). Quasi-periodic oscillations (QPOs) are routinely seen in XRB PSDs. 
They are generally divided into two broad classes, low- and high-frequency QPOs, which 
usually appear in the ranges $\sim$ 0.05 -- 30 Hz and $\gtrsim$100 Hz, respectively.

Broadband X-ray PSDs have also been constructed for AGNs, allowing modeling of the 
overall PSD shape and also showing the red-noise variability at high frequencies.
However, the quality of the data precludes complex modeling of the PSD shape; for 
example, the QPOs routinely seen in XRBs, even if they existed in AGN, could not likely be 
detected with currently available AGN data (Vaughan \& Uttley 2005). However, modeling 
AGN PSDs with broken power-law models has successfully yielded breaks on time scales 
of a few days or less (Uttley, M$^{\rm c}$Hardy \& Papadakis 2002; Markowitz \et\ 2003b 
[hereafter M03]; Marshall \et\ 2004; M$^{\rm c}$Hardy \et\ 2004), with PSD power-law 
slopes $\sim$--2 above a break and $\sim$--1 below it. The similarity in Seyferts' and 
XRBs' broadband PSD shapes and the suggested possible scaling of the PSD break time scale with 
black hole mass (e.g., M03, M$^{\rm c}$Hardy \et\ 2004) support the suggestion of 
similar X-ray variability mechanisms being present in both classes of accreting compact 
objects. The picture emerging is one in which Seyferts and XRBs are scaled versions of 
each other in black hole mass and X-ray variability time scale, though it is not yet 
clear if Seyferts are direct analogues of XRBs in the low/hard or high/soft state 
(see e.g., M$^{\rm c}$Hardy \et\ 2004, Markowitz \& Uttley 2005). An additional link 
between Seyferts and XRBs arises in the linear relation between flux and absolute 
rms variability (e.g., Uttley \& M$^{\rm c}$Hardy 2001, Edelson \et\ 2002) that 
exists in both AGN and XRBs.

The observed cross-spectral properties, namely the coherence function (Vaughan 
\& Nowak 1997) and phase lags, are also similar in both object classes, further 
supporting this notion. The coherence is generally seen to be unity over a wide range
of temporal frequencies in both AGNs and XRBs, though with a deviation from unity
above the high-frequency PSD break frequency (e.g., Nowak \et\ 1999a, 1999b; 
Papadakis, Nandra \& Kazanas 2001, Vaughan, Fabian \& Nandra 2003b [hereafter 
VFN03], M$^{\rm c}$Hardy \et\ 2004). Relatively harder X-rays are seen to generally 
lag those at softer X-ray energies, with the lag between bands increasing with 
increasing energy separation; this is consistent with a basic prediction of the 
Comptonization hypothesis. However, the lags are also observed to depend on temporal
frequency: more slowly varying components show a relatively longer time lag
(e.g., Miyamoto \& Kitamoto 1989, Miyamoto \et\ 1991, Nowak \et\ 1999a in XRBs; 
Papadakis, Nandra \& Kazanas 2001, VFN03, Vaughan \& Fabian 2003, M$^{\rm c}$Hardy 
\et\ 2004 in Seyferts). Additionally, the slope of the PSD above the high-frequency 
break is observed to flatten as energy increases (e.g., Nowak \et\ 1999a, 
Lin \et\ 2000 in XRBs; Nandra \& Papadakis 2001, VFN03, M$^{\rm c}$Hardy \et\ 2004 
in Seyferts). For instance, using Fourier-resolved spectroscopy, Papadakis, 
Kazanas \& Akylas (2005) found that, above 2 keV, the ratio of hard X-ray 
variability to soft X-ray variability increased towards relatively higher 
temporal frequencies. Similarly, Leighly (2004) also found the structure function of the
narrow line Seyfert 1 galaxy 1H~0707--495 to flatten towards higher
photon energies on time scales $\lesssim$ 3 ksec.

In this paper, we investigate the relationship between flux variations in different 
energy bands in the Seyfert~1 galaxy NGC~3783. M03 established the presence of a 
break in this object's 2--10 keV PSD at 4$\times$10$^{-6}$ Hz. In addition, M03 
found tentative evidence for a second, low-frequency break near 2$\times$10$^{-7}$ Hz, 
below which the PSD power-law slope flattens to 0. This makes NGC~3783 one of only 
two Seyfert galaxies (along with Ark~564; M03) known so far in which dominance of 
a $f^{-1}$ power-law component at the lowest temporal frequencies can be ruled out. 
M03 did not explore the energy dependence of the PSD slope above the break.
In this work, we study the cross-spectral properties and energy-dependence 
of the high-frequency PSD of NGC~3783 using data obtained from an \xmm\ long-look 
and from regular monitoring by the {\it Rossi X-ray Timing Explorer} ({\it RXTE}).
$\S$2 describes the \xmm\ and \xte\ light curve extraction. Softness ratios, 
high-frequency PSD measurement, cross- and auto-correlation functions, and
cross-spectral analysis are presented in $\S$3. $\S$4 compares these results 
to previous findings for AGN and XRBs and discusses these results in the context of
Comptonization models and X-ray variability mechanism models. 

\section{Observations and Data Reduction}

\subsection{{\it XMM-Newton} Data Reduction}

NGC~3783 was observed by \xmm\ on 2001 Dec.\ 17--21, over two full satellite 
orbits, revolutions 371 and 372. This sampling is hereafter referred to 
as short-term sampling. This paper uses data taken with the European 
Photon Imaging Camera (EPIC), which consists of one pn CCD back-illuminated 
array sensitive to 0.15--15 keV photons (Struder \et\ 2001), and two MOS 
CCD front-illuminated arrays sensitive to 0.15-12 keV photons (MOS1 and 
MOS2, Turner \et\ 2001). Data from the pn were taken in Small Window Mode; 
data from the MOS detectors were taken in Full Window Mode. This paper 
will concentrate primarily on data obtained with the pn, though MOS2 
data were also analyzed for consistency. Data from the MOS1 were taken 
in Fast Uncompressed mode and are not used here. 
The medium filter was used for all detectors. Light curves were extracted 
using XMM-SAS version 6.0 using standard extraction procedures. Data were 
selected using event patterns 0--4 for the pn and 0--12 for the MOS2. Hot, 
flickering, or bad pixels were excluded. Source light curves were extracted 
from a circular region of radius 40$\arcsec$; background light curves were 
extracted from circles of identical size, centered $\sim$3$\arcmin$ away 
from the core. Due to increased background levels at the end of each 
satellite orbit, the final 7 ksec and 13 ksec were excluded from the ends 
of revolutions 371 and 372, respectively. This reduction yielded two pn 
light curves of 129.9 ksec (rev.\ 371) and 123.8 ksec (rev.\ 372) in 
duration, separated by a 42.4 ks gap. The MOS2 light curves were 130.5 
ksec (rev.\ 371) and 121.2 (rev.\ 372) in duration, separated by a 41.7 
ksec gap. The revolution 371 light curves were uninterrupted except for
gaps of 4.8 ksec (pn) and 5.0 ksec (MOS2) that occurred about 57 ksec
after the start of the observation. The revolution 372 light curves were 
uninterrupted. Data were extracted over the 0.2--12 keV (total), 
0.2--0.7 keV (soft), 0.7--2 keV (medium), and 2--12 keV (hard) bandpasses
(also referred to as T, S, M and H bands, respectively, below). 
The hard bandpass was further subdivided into the 2--4, 4--7, and 7--12 keV 
bandpasses (hereafter referred to as the H1, H2, and H3 bands, respectively)
for purposes of calculating the coherence function below. The 0.2--12 keV pn
light curves for each revolution, binned to 1200 sec, are shown in Figure~1.  
Mean count rates for all of the pn light curves are listed in Table 1. 
Count rates and signal-to-noise values for the MOS2 light curves are not 
listed, but are usually $\sim$1/2 of the corresponding pn light curve values. 

%%%%%125004957.  125062112  ed  pn  57.2    
%%%%%      gap                      4.8
%%%%%125066904  125134884  ee   pn  68.0      total371 = 129.9
%%%%%  gap                           42.4 
%%%%%125177222  125300997.67  ef pn     123.8

%%%%%125004327.49 -  125061987  mos2  er 57.7
%%%%%      gap                            5.0
%%%%%125066996 - 125134856.02  mos2 es   67.9 ksec  total 371 = 130.5
%%%%%      gap                      41.7  
%%%%%125176529 - 125300999.7  mos2  et   124.5

%%%%%m1  DATAMODE= 'IMAGING '          / Instrument mode
%%%%%m1 SUBMODE = 'FastUncompressed'   / Guest Observer mode
%%%%%m1 CCDMODE = 'PrimeFullWindow'    / CCD Mode

%%%%%pn DATAMODE= 'IMAGING '           / Instrument mode (IMAGING, TIMING, BURST, etc.)
%%%%%pn SUBMODE = 'PrimeSmallWindow'   / Guest Observer mode
%%%%%pn CCDMODE =                    3 

%%%%%m2 DATAMODE= 'IMAGING '           / Instrument mode (IMAGING or TIMING)
%%%%%m2 SUBMODE = 'PrimeFullWindow' 
%%%%%m2 CCDMODE = 'PrimeFullWindow'    

\subsection{\xte\ Data Reduction}

NGC~3783 was monitored by \xte\ once every 4.3 d from 1999 Jan.\ 02 -- 
2002 Feb.\ 06 (hereafter referred to as long-term sampling) and once every 
other orbit (3.2 hr) from 2001 Feb.\ 20 -- 2001 Mar.\ 12 (medium-term sampling); 
see M03 for observing details. Data were taken using {\it RXTE}'s proportional 
counter array (PCA; Swank 1998), which consists of five identical collimated 
proportional counter units (PCUs). For simplicity, data were collected
only from those PCUs which did not suffer from repeated breakdown during 
on-source time (PCUs 0 and 2 before 2000~May~12; PCU~2 only after 2000 May 12). 
Count rates quoted in this paper have been normalized to 1 PCU. Only PCA 
STANDARD-2 data were considered. The data were reduced using standard 
extraction methods and {\sc FTOOLS~v5.3.1} software. Data were rejected 
if they were gathered less than 10$\arcdeg$ from the Earth's limb, if
they were obtained within 30~min after the satellite's passage through 
the South Atlantic Anomaly, if {\sc ELECTRON0}~$>$~0.1 ({\sc ELECTRON2} 
after 2000~May~12), or if the satellite's pointing offset was greater 
than 0$\fdg$02.

As the PCA has no simultaneous background monitoring capability, background 
data were estimated by using {\sc pcabackest~v3.0} to generate model files 
based on the particle-induced background, SAA activity, and the diffuse X-ray 
background. The 'L7-240' background models appropriate for faint sources
were used. This background subtraction is the dominant source of systematic 
error in \xte\ AGN monitoring data (e.g., Edelson \& Nandra 1999). Counts were 
extracted only from the topmost PCU layer to maximize the signal-to-noise ratio. 
Light curves were generated over the 2--18 keV, 2--4 keV, 4--7 keV, 7--12 keV, 
and 12--18 keV bandpasses (the latter four are hereafter referred to as the H1, 
H2, H3 and H4 bands, respectively). The medium-term light curves were binned to 22.9 ksec to 
increase the variability to noise ratio; the long-term light curves were binned 
to 4.3 days. Standard errors were derived from the data in each orbital bin. The 
2--18 keV \xte\ light curves are shown in Figure~1. Mean count rates for all 
light curves are listed in Table~1.

\section{Analysis}

The variability analysis presented in this paper consisted of several tests.
Because the quality of the data and the variability-to-noise ratio were 
different for all three sampling patterns, and because the \xte\ and \xmm\ 
bandpasses only overlap at 2--12 keV, not all light curves could be used for 
each timing measurement.

Analyses using the short-term data concentrated primarily on the \xmm\ pn light 
curves. The MOS2 data were of lower signal-to-noise; they were also analyzed, 
yielding results consistent with the pn data, and they are not discussed in 
detail here. Because gaps such as the one between the rev.\ 371 and rev.\ 372 
\xmm\ light curves complicate timing analysis, the two light curves from each 
revolution were treated separately in the analyses below (as opposed to combining 
the two light curves and interpolating across the large gap). The predicted level 
of variability power due to Poisson noise was calculated as $P_{{\rm Psn}}$ = 
2($\mu$ + $B$)/ $\mu^2$, where $\mu$ and $B$ are the net source and background 
count rates, respectively, and is listed in Table 1 for each light curve. For 
non-continuously observed data such as the \xte\ data here, $P_{{\rm Psn}}$ is 
multiplied by the ratio of the mean sampling time to the duration of each 
individual visit, e.g., 22.9 for the medium-term data binned to 22.9 ksec and 
368.64 for the long-term data binned to 4.3 days. The power due to Poisson noise in 
the MOS2 data is not listed, but is typically higher by a factor of $\sim$3 
compared to the pn data.

Short-term variability amplitudes are presented in $\S$3.1; variability amplitudes 
for the long and medium time scale data have been discussed previously in M03 
and Markowitz \& Edelson (2004). Softness ratio (SR) light curves are presented in
$\S$3.2. The energy dependence of the high frequency PSD of NGC 3783 is presented 
in $\S$3.3, using the short-term data. Cross correlation functions are presented 
in $\S$3.4 for all three time scales. The cross-spectral properties, including the
interband coherence properties and temporal frequency-dependent phase lags, are 
discussed in $\S$3.5.

\subsection{Variability Amplitudes}      %%%%%Sect. 3.1

The \xmm\ light curves were rebinned to the approximate time scale at which the 
Poisson noise begins to dominate over the intrinsic red-noise variability. 
This ensured that the fractional variability amplitude was quantified over 
temporal frequencies where the intrinsic red-noise variability is greater 
than the variability due to Poisson noise. This time scale was about 1200 s 
for the pn and 1600 s for the MOS2. Fractional variability amplitudes $\fv$ 
and uncertainties were calculated using the formulation of e.g., Vaughan \et\ 
(2003a) and are listed in Table 1 for all time scales and bandpasses. For the 
short-term data, only the pn results are listed; the MOS2 data yielded virtually 
identical results, though uncertainties on $\fv$ were usually a factor of $\sim$2--3
higher compared to the pn data. For all data and time scales, $\fv$ tends to decrease 
towards higher energies, consistent with numerous previous results (e.g., Markowitz 
\& Edelson 2004 and references therein). The 2--12 keV short-term $\fv$ values 
are quite similar to those measured over similar time scales by M03 for 2--10 
keV \chandra\ light curves. 

\subsection{Spectral Variability}       %%%%% sect 3.2        
 
To explore spectral variability, softness ratio (SR) light curves were calculated 
for all time scales. The S/M, S/H and M/H ratio light curves were calculated for 
the short-term \xmm\ data binned to 1200 s, and are overplotted on the 
corresponding sum of the two sub-band light curves (e.g., S-band + M-band) 
in Figure 2. The ratios of H1/H2,  H1/H3, and H1/H4 for the long-term and 
medium-term \xte\ data are also overplotted against their summed light 
curves in Figure 2; the ratios of H2/H3, H2/H4, and H3/H4 were calculated 
for completeness but are not shown in Figure 2. In these plots, all SR 
light curves (solid lines) and summed light curves (filled circles) 
have been mean-subtracted and normalized by the standard deviation to 
allow us to directly visually compare variations in flux or softness ratio.
In general, the SR tracks the overall flux, consistent with numerous previous 
observations (e.g., Papadakis \et\ 2002), but there are tentative 
indications in 
the short-term plots for spectral softening to precede increases in the 
summed flux by very roughly 5--10 ksec. In $\S$3.4, we quantify this lag using 
cross-correlation functions; subsequent Monte Carlo simulations that quantify 
the significance of this claim show that it is preliminary at best. 
On medium and long time scales, there is no obvious lag.

\subsection{PSD energy dependence}    %%%%%sect 3.3

Previous studies (e.g., Nowak \et\ 1999a, VFN03, M$^{\rm c}$Hardy \et\ 2004)
indicate that in Seyferts and XRBs, the modeled PSD high-frequency slope 
generally flattens as energy increases, though break frequency generally 
does not depend on photon energy. The amplitude of the PSD at the break frequency 
generally decreases as photon energy increases, leading to light curve 
variability amplitudes which generally decrease as photon energy increases. 
The high-frequency PSD break for NGC~3783 measured by  M03, 4$\times$10$^{-6}$ Hz, 
falls in the temporal frequency range covered by the medium-term \xte\ data. However, 
as \xte\ cannot detect $<$2 keV photons, the present \xte\ data are not 
adequate to test these notions over a sufficiently wide energy range. 
Here, we assume a constant PSD break frequency of 4$\times$10$^{-6}$ Hz 
for all photon energies, and instead concentrate on the relative slopes 
of the high-frequency PSD at soft, medium, and hard energies using 
only the \xmm\ pn data. PSDs derived from the MOS2 data are not 
presented here; the lower count rates and higher levels of power due to
Poisson noise mean the MOS2 is of much lower quality, but the
MOS2 results were generally consistent with those obtained using the pn data.

Power spectra for the soft, medium, and hard bandpasses were constructed 
in a manner similar to Uttley, M$^{\rm c}$Hardy \& Papadakis (2002), 
M03 and VFN03. The revolution 371 light curves were linearly interpolated 
across missing data points (the percentage of missing data was small, 4$\%$; 
no data was missing for revolution 372). Each light curve's mean was 
subtracted. The power spectra were constructed using a Discrete Fourier 
Transform (DFT; e.g., Oppenheim \& Shafer 1975). The periodogram was 
constructed from the DFT using the normalization of van der Klis (1997).
Periodograms for the revolution 371 and 372 data were constructed 
separately (e.g., the two orbits were not combined into a single 
light curve).

Following Papadakis \& Lawrence (1993), each periodogram was binned 
every factor of 1.4 in temporal frequency (0.15 in the logarithm), 
with the lowest three frequency bins widened to accommodate a minimum 
of two periodogram points. The resulting PSDs spanned 
1.10$\times$10$^{-5}$ Hz -- 3.39$\times$10$^{-4}$ Hz and 
1.15$\times$10$^{-5}$ Hz -- 3.39$\times$10$^{-4}$ Hz
for revolutions 371 and 372, respectively, factors of $\sim$1.5 
orders of magnitude in dynamic range. 

We note that for light curves binned to 200 s, the resulting
periodograms were dominated at the highest temporal frequencies 
(above $\sim$4$\times$10$^{-4}$ Hz) by white noise power due to 
Poisson noise, confirming the need to rebin the light curves to 1200 s.
For all periodograms, the observed white noise power level was equal to the
predicted level of power due to Poisson noise (listed in Table 1).

Two methods were used to probe the energy dependence of the
high-frequency PSDs. First, the Monte Carlo method of Uttley, 
M$^{\rm c}$Hardy \& Papadakis (2002) was used to quantify
the intrinsic, underlying PSD slope, hereafter referred to 
as --$\beta$. Second, a Monte Carlo method that uses 
a least-squares fit to the observed (not underlying) PSD 
was used to probe relative high-frequency slope differences.

\subsubsection{The Underlying PSD slope}      %%%%% sect 3.3.1

The observed, binned PSDs had poorly defined errors due to 
the low number of periodogram points in each PSD bin
(e.g., Papadakis \& Lawrence 1993). Furthermore, the PSDs 
suffered from the systematic distortion effect of red-noise leak,
whereby variability trends on time scales longer than the observed 
duration can contribute additional variance to the observed light curve
and contribute extra, ``leaked'' variability power to the observed 
PSD (e.g., van der Klis 1997). The PSD distortion effect of aliasing 
is a frequent issue in PSD measurement; with aliasing, discrete, 
non-continuous sampling fails to distinguish variability trends on 
time scales longer and shorter than the sampling time and 
effectively adds extra variability power to the observed PSD.
However, aliasing was not expected to be a factor here because 
these PSDs were derived from continuously sampled data.

We employed the Monte Carlo fitting procedure of Uttley, M$^{\rm c}$Hardy \&
Papadakis (2002) to get rid of distortion effects, properly assign 
errors, and quantify the underlying PSD shape. In this procedure, a test 
model PSD shape is assumed, and a large number ($N$, at least 100) of light 
curves are simulated using the method of Timmer \& K\"{o}nig (1995). The 
light curves are simulated with a duration much longer than the observed 
duration to ensure that the light curves account for variability on time 
scales longer than the observed duration (that is, they contain a variability 
contribution due to red noise leak). They are also simulated with a time 
resolution higher than the observed sampling time. The simulated light 
curves are then resampled to match the observed light curves to ensure 
that the effects of sampling are identical for both the observed and 
simulated light curves. The $N$ PSDs are measured from the simulated data and 
averaged to form an ``average distorted model'' PSD with well-determined
errors, and so-called because they suffer from the aforementioned 
distortion effects. Finally, the observed PSD is compared to the 
distribution of the $N$ PSDs to estimate the confidence for the null 
hypothesis that the specified PSD model shape is correct. For further 
details of this analysis see $\S$3--4 of Uttley, M$^{\rm c}$Hardy \& 
Papadakis (2002) or $\S$3 of M03.

For the present analysis, the PSD model shape tested was an unbroken 
power law of the form \[ P(f) = A (f/f_0)^{-\beta} + P_{{\rm Psn}} \] 
where the normalization $A$ is the PSD amplitude at the break frequency $f_0$, %%%% = 4.0$\times$10$^{-6}$ Hz, 
$\beta$ is the power law slope, and $P_{{\rm Psn}}$ is the constant 
level of power due to Poisson noise. The limited dynamic range of 
the data precludes testing more complicated PSD model shapes. 
$P_{{\rm Psn}}$ was added to the model as opposed to being subtracted 
from the data, to avoid the possibility of obtaining unphysical 
negative powers.

The range of $\beta$ tested was 1.00--4.00 in increments of 0.05. The
break frequency was fixed at 4.0$\times$10$^{-6}$ Hz in all cases, 
though the normalization $A$ was allowed to vary. Light curves 
were simulated with a duration 250 times the observed duration, to 
account for red noise leak. The light curves were simulated with a 
time resolution of 200 s, then binned to 1200 s (1600 s for the MOS2 data).
The light curve durations were trimmed to 129.9 ksec or 123.8 
ksec for revolutions 371 and 372, respectively. The simulated 
light curves were resampled to match the observed light curves:
4 data points in the revolution 371 data from 57.0 ksec to 61.8 
ksec after the start of the observation were destroyed and 
linearly interpolated from adjacent points. Of the simulated PSDs, 250 
were used to determine the the ``average distorted model'' PSD with 
well-defined errors, and 10$^3$ randomly selected sets 
of PSDs were used to probe the simulated $\chi^2$ distribution. 
Because the power due to Poisson noise differs slightly between 
the two revolutions, the PSD model shape was fitted separately 
to the observed PSD for each revolution, with the respective 
$\chi^2$ values from each revolution added to form a total $\chi^2$ 
used in determining the goodness of fit.

The best-fit values of $\beta$ and $A$, along with the corresponding 
likelihood of acceptance (defined as 1 -- $R$, where $R$ is the 
rejection confidence), are listed in Table~2. Following M03, the 
uncertainties on $\beta$ correspond to values 1$\sigma$ above the
likelihood of acceptance for the best-fit value on a Gaussian
probability distribution (i.e., the amount $\beta$ has to change 
for the fit to be less likely by 1$\sigma$). The errors on $A$ were 
determined from the rms spread of 10$^3$ randomly selected sets 
of simulated PSDs. The best-fit PSD model shapes are illustrated 
in Figure~3. The hard band results are consistent with 
those presented in M03 for the high-frequency PSD shape 
derived from 2--10 keV \chandra\ observations.

Due to the limited dynamic range of these PSDs, the Monte Carlo 
procedure yields very poor constraints on both the high-frequency 
PSD slope $\beta$ and the amplitude $A$. Furthermore, upper limits 
on $\beta$ cannot be constrained, for two reasons. First, the presence 
of a constant Poisson power level dominates the steepest PSD slopes 
tested, leading to a degeneracy such that most values of 
$\beta$$\gtrsim$3 yield similar rejection probabilities. Second,
excessive red-noise leak from very steep PSD slopes increases the PSD
errors, with the result that such large errors decrease the reliability of
the values of the rejection probability (e.g., M03). Formally, 
then, we have only lower limits to $\beta$. However, looking at 
the best-fit results, we cannot rule out consistency with 
previous results that showed that the PSD slope flattens as 
energy increases. Although the PSD amplitude $A$ at the break is 
also poorly constrained, the best-fit values are not inconsistent 
with the amplitude decreasing as photon energy increases.

\subsubsection{Relative differences in PSD slope}         %%%%% sect 3.3.2

Complementary to finding constraints on the intrinsic slope $\beta$ 
using the Monte Carlo procedure above, a power-law model was directly 
fit to the observed, ``raw'' PSDs. This was done 
to obtain additional constraints on $\beta$ as a function of energy, 
since the Monte Carlo routine above was not able to provide tight 
constraints on $\beta$ given the limited dynamic range of the PSDs.
This procedure did not aim to directly identify the underlying 
high-frequency PSD slope $\beta$, but rather to qualitatively 
identify PSD power-law slopes relative to each other in the 
presence of red noise leak.

%%%%The observed PSDs suffer from red-noise leak, but because 
%%%%the sampling in each energy bandpass was identical, the sampling 
%%%%window functions in temporal frequency space were identical. 
%%%%With roughly similar PSD shapes from one bandpass to the next, 
%%%%the expected level of red-noise leak present in each PSD is roughly 
%%%%similar for each bandpass and for both revolutions. 

We used a power-law model shape of the same form as in the Monte Carlo 
analysis, \[ P(f) = A (f/f_0)^{-\beta_{{\rm obs}}} + P_{{\rm Psn}} \] 
where $\beta_{{\rm obs}}$ refers to the ``raw,'' observed PSD slope, 
which is systematically affected by red noise leak. The effect of 
red-noise leak is to add an $f^{-2}$ noise component to a PSD, thus 
flattening any intrinsic PSD with power-law slope steeper than --2 
(van der Klis 1997). We fit using a ``least-squares'' routine, specifically, 
the ``GRIDLS'' procedure of Bevington (1969), and using the PSD 
errors assigned from the best-fit model PSDs constructed
during the Monte Carlo procedure above. Because $P_{{\rm Psn}}$ 
differed slightly between the two revolutions, the revolution 
371 and 372 PSDs were again fit separately for each revolution,
with the respective $\chi^2$ values from each revolution added
to form a total $\chi^2$ used in determining the goodness of fit.
The best fitting slopes and normalizations are summarized in Table~3. 
However, values of $\chi^2$ are very low due to the large PSD error sizes
and the limited dynamic range. The best-fit values for $\beta_{{\rm obs}}$ 
tend to flatten as energy increases, consistent with
the results of the Monte Carlo fitting above, but
the reader is reminded that due to the presence of red-noise leak,
these slopes are NOT direct estimates of the underlying PSD model parameters.
This contrasts with the results from the Monte Carlo analysis (Table~2 and
Figure~3), which show best-fit estimates to the underlying, intrinsic 
PSD model shape.

The exact amount of power ``leaked'' from below $f_{{\rm min}}$ (defined as 
1/$D$, where $D$ is the observation duration) to above $f_{{\rm min}}$
is a stochastic quantity, but the amount of red noise leak power 
present in the observed PSDs can be estimated by integrating the 
underlying, intrinsic PSD model shape. We used the best-fit 
doubly-broken PSD model of NGC 3783 from M03,
\[P(f)= \left\{ \begin{array}{ll}
                              A_l,          & f \le f_l \\
                              A(f/f_h)^{-1},  & f_l < f \le f_h \\
                              A(f/f_h)^{- \beta},  & f > f_h \end{array}  
\right. \]
where $A_l$ is the PSD amplitude below the low-frequency break 
$f_l$ = 2$\times$10$^{-7}$ Hz, where the PSD has zero slope. The 
quantity $A$~=~A$_l$$(f_h/f_l)^{-1}$ is the 
PSD amplitude at the high frequency break $f_h$ = 4$\times$10$^{-6}$ Hz. 
Values of $\beta$ and $A$ were the slopes and power amplitudes
from the best-fit Monte Carlo results in $\S$3.3.1. 
We assumed that the only PSD model parameters
that vary between bandpasses are $\beta$ and $A$.
We also assumed that the PSD of NGC~3783 does not exhibit 
non-stationarity between any of the
observations, a hypothesis tested in $\S$3.3.3 below. 
Integration over temporal frequencies from 4$\times$10$^{-8}$ Hz (the temporal
frequency corresponding to 250 times the duration of the pn light curves)
to 1$\times$10$^{-5}$ Hz (the minimum temporal frequency sampled here)
showed that the predicted power
due to red noise leak is highest in the soft band and lowest in the hard
band, with a factor of 3.2 greater in the soft band compared to
the hard band. The relatively softer
PSD slopes have thus been flattened to a greater degree relative to harder bands;
given that the $\beta_{{\rm obs}}$ steepen toward higher 
energies, the intrinsic PSD slopes must also steepen towards higher energies.

Finally, the differences in the best-fit $\beta_{{\rm obs}}$ between the 
soft, medium and hard bands were calculated:
$\Delta$$\beta_{{\rm obs,S-M}}$ = 0.31,
$\Delta$$\beta_{{\rm obs,S-H}}$ = 0.49, and
$\Delta$$\beta_{{\rm obs,M-H}}$ = 0.19,
where e.g., $\Delta$$\beta_{{\rm obs,S-H}}$ equals $\vert$ $\beta_{{\rm obs}}$ $\vert$ 
for the soft band minus $\vert$ $\beta_{{\rm obs}}$ $\vert$ for the hard band.
A Monte Carlo simulation procedure was done to assess the significance
of each of these three slope differences.
This procedure tested the null hypothesis of simultaneous
light curves in two different bandpasses intrinsically having identical 
underlying PSD shapes and unity coherence (see $\S$3.5 for a definition of
coherence). Rejection of this null hypothesis 
at high confidence would mean that in the limit of unity coherence,
the two light curves are derived from underlying PSDs with different shapes.
In this Monte Carlo procedure, for each pair of relatively soft and hard 
light curves, underlying PSD shapes were assumed:  
the model PSD shape used was the same as in the Monte Carlo 
procedure in $\S$.3.3.1.  Identical power-law slopes were assumed for both
light curves; values of $\beta$ from 1.0--4.0 were 
tested in increments of 0.1. For each revolution, 200 light curves 
were simulated for both the relatively soft and hard bands
using the same random number seed to ensure unity coherence between them.
Light curves were simulated with a duration 250 times the observed duration
in each revolution to adequately ensure the presence of long-term variability
trends associated with red noise leak.
The time resolution for the simulations was 200 s.
The soft and hard light curves were clipped to the appropriate duration 
(129.9 ks and 123.8 ksec for revolutions 371 and 372, respectively)
and binned to 1200 s. The simulated light curves were 
resampled to match the observed light curve (four flux bins
were removed from the revolution 371 soft and hard simulated light curves at 
57.0--61.8 ksec after the start of the observation
and linearly interpolated). The light curves were rescaled to match the 
mean count rate and excess variance of the observed
light curves. This rescaling accounts for the
difference in the expected amount of red noise power leakage between the 
different bandpasses. PSDs were measured in the same manner as the 
observed data. PSD errors for the simulations were 
taken from the best-fit corresponding average distorted model 
from the Monte Carlo procedure described in $\S$3.3.1.
To simulate the effect of Poisson noise in the simulated light curves, 
each flux point was randomized according to the Poisson distribution.
Least-squares fitting was done in the same manner as the observed PSDs,
using the same model shape, a red-noise power law plus a white noise 
component due to Poisson noise. The difference in resulting slopes,  
e.g., $\Delta$$\beta_{{\rm sim, E1-E2}}$ was calculated, where E1 and E2
represent the relatively softer and harder light curves, respectively.
A distribution of $\Delta$$\beta_{{\rm sim, E1-E2}}$ was formed from
the 200 sets of simulations for each test value of $\beta$; the distribution
arose due to stochastic differences in measured PSD slope from one simulated light curve
to the next and due to the minor systematic effects of Poisson noise, but
could not be due to differences in intrinsic PSD slope.
The $\Delta$$\beta_{{\rm sim, E1-E2}}$ distribution was then compared
against the observed values of $\Delta$$\beta_{{\rm obs}}$.
The observed value of $\Delta$$\beta_{{\rm obs,S-M}}$ was
never reached more than twice in any of the 200 simulations
for any initial value of $\beta$. Similarly, the observed value
of $\Delta$$\beta_{{\rm obs,S-H}}$ was never reached more than 
once out of the 200 simulations for any initial value of $\beta$, 
and $\Delta$$\beta_{{\rm obs,M-H}}$ was never reached more than
3 times out of 200 simulations.
We can thus reject at 99.5$\%$ confidence the null hypothesis of both 
the S-band and M-band light curves having identical intrinsic 
PSD shapes in the limit of unity coherence;
corresponding rejection confidences for the soft--hard
and medium--hard combinations are $>$99.5$\%$ and 98.5$\%$, respectively.

\subsubsection{Stationarity tests}      %%%%% sect 3.3.3

We tested a fundamental assumption inherent in AGN PSD analysis, 
that the variability process does not display non-stationarity 
over long time scales. Following previous definitions, a 
non-stationary variability process is one whose underlying PSD 
changes with time. For instance, the PSDs of XRBs change on 
time scales of hours to days (e.g., Axelsson \et\ 2005), but if 
variability characteristics scale linearly with black hole mass,
then one would not expect AGN PSDs to significantly change shape 
on time scales less than at least decades to centuries. 

Compact accreting black holes display ``weakly stationary'' 
behavior in that the mean and variance both show scatter over 
time, although both the expectation value of the fractional variability 
amplitude and the underlying PSD shape is expected to remain constant 
in time (e.g, over time scales shorter than decades). At a given 
temporal frequency, periodogram values are scattered about the 
underlying PSD following a $\chi^2$ distribution with two degrees 
of freedom (e.g., Priestley 1981). For two PSDs that sample the same 
stationary process and are observed at two different times, 
the only difference in the PSD values at a given temporal 
frequency will be due to this scatter. Papadakis \& Lawrence (1995) 
outlined a method to test for non-stationarity (change in 
underlying PSD shape), defining a statistic $S$ to quantify 
differences between two PSDs; for a stationary process, $S$ will 
be distributed with a mean of 0 and a variance equal to 1.

We tested for consistency in PSD shape between the two \xmm\ pn 
hard band light curves and the five 170-ksec long \chandra\ hard 
band light curves obtained in 2001 February--June (see M03 and 
Kaspi \et\ 2002 for observing and data reduction details).
All light curves were trimmed to a common duration (123.8 ksec, 
the duration of the rev.\ 372 light curve) and rebinned to a 
common sampling time of 2000 s. PSDs were measured using the same 
normalization as above, and the $S$-statistic of Papadakis \& 
Lawrence (1995) was calculated. Comparing the revolution 371 and 
372 pn PSDs yielded $S$ = --0.79. Comparing the revolution 371 PSD 
to each of the five \chandra\ PSDs yielded $S$ values between 
--1.50 and --0.58; comparing the revolution 372 PSD to each of the 
five \chandra\ PSDs yielded $S$ values between --0.71 and +0.21.
Finally, comparing each of the \chandra\ PSDs in a pairwise fashion 
yielded $S$ values between --0.92 and +0.78. All of these values 
are within 2$\sigma$ of the expected values assuming stationarity 
in the data; the assumption of stationarity of the high-frequency 
power spectrum over time scales of up to ten months is thus a 
reasonable one.

\subsection{Cross-Correlation Functions}      %%%%% sect 3.4

We next measured the cross-correlation functions (CCFs) to search for
intra-X-ray lags. We compared the soft, medium and hard bands in the 
\xmm\ data, and also compared the H1, H2, H3 and H4 
bands to each other in the medium- and long-term data.
We used the Interpolated Correlation Function (ICF; 
White \& Peterson 1994), with uncertainties on the ICF peak 
lag determined using the bootstrap method of Peterson \et\ (1998). 
We also measured the Discrete Correlation Function (DCF; 
Edelson \& Krolik 1988); results were found to be virtually 
consistent with those obtained via the ICF.

For the short-term data, we used the pn light curves binned to 
1200 s (1600 s for the MOS2). Again, only the results for the pn 
data are presented; the results for the MOS2 data were consistent.
%%%%% For the medium-term \xte\ data, the light curves were binned to 22.9 ksec. %%%%% already said this above
DCF and ICF results are summarized in Table~4. ICF results
are plotted in Figure~4; the (virtually identical) DCF results are not plotted.
Positive lags are defined as relatively harder band 
variations being delayed with respect to the softer band.

For the short-term data, all peak lags $\tau$
are consistent with delays between 0 and $\lesssim$+2 ksec.
Conservatively, one can assign a 2$\sigma$ limit of twice the mean
sampling bin, or $\vert \tau \vert \lesssim$2.4 ksec.
The revolution 371 and 372 cross-correlation functions 
are not identical in profile. However, this difference is not
sufficient for any claim of non-stationarity, at least for the
hard band, given the results of $\S$3.3.3 above.
The cross correlation results reveal an asymmetry
towards hard lags at short time scales, especially evident in the
revolution 371 data. Such an asymmetry has been seen previously in
XRBs (Maccarone \et\ 2000) and AGN (M$^{\rm c}$Hardy \et\ 2004), and is
indicative of something more complex than a simple lag that is 
constant in temporal frequency. 
There are hints only of positive lags, but moreso in the revolution
371 data, and one cannot definitely conclude that significant lags exist
in either short-term data set. For the medium-term and long-term results,
peak lags are all consistent with zero; there is no evidence for
peak lags outside the (conservative) 2$\sigma$ lag limits
of $\vert \tau \vert \lesssim$ 45.8 ksec (medium-term) or
8.5 d (long-term).

For all time scales and bandpasses, the peak correlation coefficient
$r_0$ tends to decrease as the relative energy separation between 
the two bands increases. Monte Carlo simulations were required to assess 
the significance of this result and distinguish whether it is intrinsic 
to the source or an artifact due to Poisson noise. The null hypothesis 
tested was that the two light curves have unity coherence and peak correlation 
coefficient of 1, with any lack of perfect correlation being due solely 
to Poisson noise. In these simulations, for each pair of relatively soft 
and hard light curves, a PSD model shape was assumed. For the short-term 
data, the best-fit PSD slope and amplitude were taken from the 
Monte Carlo results of $\S$3.3.1. For the medium- and long-term data,
the best fitting doubly-broken PSD model shape of M03 was used. Of the simulated light curves, 200 were 
simulated for both the relatively soft and hard bands using the 
same random number seed to ensure unity coherence between them. Light curves 
were initially generated with a duration 250 times the observed
duration. The initial simulated time resolution was 200 s, 2.29 ksec, or
0.43 days for the short-, medium-, or long-term data, respectively; 
light curves were clipped to the observed duration and resampled to match 
the sampling of the observed light curves. The light curves were 
rescaled to match the mean count rate and excess variance of the observed
light curves. To simulate the effect of Poisson noise in the simulated 
light curves, each flux point was randomized according to the Poisson 
distribution. Cross-correlation peak coefficients $r_0$ were measured 
using the ICF as above; these simulated $r_0$ will be less than 1 due 
only to the presence of Poisson noise. For each of the 200 pairs of 
light curves, a distribution of (simulated) $r_0$ was formed and compared 
to the observed value of $r_0$. For all pairs of light curves and time 
scales, the fraction of simulated values of $r_0$ exceeded by the observed 
value of $r_0$ is listed in Table 4. For the revolution 371 short-term data, 
the null hypothesis that the two light curves have unity coherence and 
peak correlation coefficient of 1 is rejected at 91.5$\%$ confidence for 
the soft--hard CCF, and $>$99.5$\%$ confidence for the medium--hard CCF, 
but cannot be rejected significantly for the soft--medium CCF. For the 
revolution 372 short-term data, the null hypothesis is rejected at 94.5$\%$ 
confidence for the soft--medium CCF, 97.0$\%$ confidence for the soft--hard 
CCF, and 99.5$\%$ confidence for the medium--hard CCF. For the 
medium-term data, the null hypothesis is rejected at $>$95.5$\%$ 
confidence for the H1--H2, H1--H3, and H2--H3 CCFs. However, the 
null hypothesis cannot be rejected at high significance for the H1--H4, 
H2--H4 and H3--H4 CCFs, due to the lower signal-to-noise in the H4 band. 
A repeat of this analysis with much larger time bins (e.g., 46 ksec 
or 92 ksec time bins) would further raise the variability-to-noise 
ratio of the light curve but would not leave enough data for an 
adequate CCF. For the long-term data, the null hypothesis is rejected 
at $>$99.5$\%$ confidence for all CCFs. Overall, there is thus a 
general consensus that the peak correlation coefficient decreases as the
energy separation of the pairs increases.

Autocorrelation functions (ACFs) for the short-term light curves were also 
calculated using the ICF and DCF; results for the ICF are plotted 
in Figure 5. The short-term ACFs show a tendency for the central 
peak to become relatively more narrow at harder energies. This is 
consistent with the flattening of PSD slope with increasing photon 
energy as described in $\S$3.3. A similar phenomenon was noted by 
Maccarone \et\ (2000) for Cyg X-1 in the low/hard state.

This energy dependence of the ACF and the high-frequency PSD
slope suggest that for a given variability event in Seyfert galaxies or XRBs, the relatively 
harder X-ray band is associated with shorter rise and decay times. That 
is, for a phenomenological variability event consisting solely of
a rapid flux increase followed by a decrease, the rise and decay times
will be shorter for the relatively harder band. However, the sub-band 
CCF analysis (Figure 4) indicates that the maxima of such events should
coincide (to within $\sim$ 2 ksec in the case of NGC~3783). Therefore, 
one might expect the rise to appear in the softer band first, 
leading to an increase in the softness ratio (SR), followed by the 
harder band rising a short time later.

To attempt to provide support for this picure in NGC 3783, CCFs 
were calculated to search for lags 
between the SR light curves and the summed flux light curves,
e.g., (S/H) versus (S + H). DCF and ICF results are listed in Table~5 
and shown in Figure 6, using the same time resolutions as for the 
earlier CCFs; positive lags are defined as changes in the SR light 
curve leading changes in the summed light curve. For the short-term data, 
all peak lags are consistent with delays between $\sim$0 and $\sim$+8 ksec, 
including delays larger than the conservative 2$\sigma$ limit of 2.4 ksec, 
e.g., in revolution 371. Formally, the revolution 372 lags are all 
consistent with zero. For the medium- and long-term data, peak lags 
are all consistent with zero, with no evidence for lags outside the 
2$\sigma$ limits of 45.8 ksec or 8.5 days, respectively.

One danger inherent in cross-correlation analysis of red-noise light curves 
is that a small number of large amplitude events or flux changes can
dominate the CCF function. For example, in revolution 371, the positive 
lag may be driven primarily by the big rises in SR and the summed flux 
at $\sim$ 50--90 ksec after the start of the observation.
In revolution 372, the dip and rise in the SR light curves 
$\sim$30 ksec after the start of the observation seems to precede 
a similar dip and rise about 10 ksec later in the summed light curves.
It was therefore necessary to perform Monte Carlo simulations to 
assess the significance of the SR--summed band lags in the short-term data, 
the null hypothesis being that the SR light curve and the summed 
light curves have zero intrinsic lag, and are derived from a pair of 
relatively soft and hard light curves with unity coherence.

The relatively soft and hard light curves were derived in a manner
identical to the CCF simulations above, with the PSD model shapes 
the same as in $\S$3.3.1, light curve pairs simulated with the same 
random number seed, and Poisson noise added to the simulated light 
curves. For each of the 200 pairs of simulated SR and summed light 
curves, a distribution of simulated lags was formed for comparison 
with the observed lags. The fraction of simulated positive lags 
greater than the observed peak lags and the observed lower limits 
to lags are listed in Table 5.

For revolution 371, the observed peak lag and lower lag limit to the
M/H SR versus summed CCF are exceeded by 4/200 and 18/200
simulated lags, respectively, so the null hypothesis is rejected at 
$>$90$\%$ confidence. For the S/M and S/H ratios in 
revolution 371, the lower limit to the observed lag is exceeded by 
at least 40 of 200 simulated lags, so the null hypothesis cannot be 
rejected at high confidence. For revolution 372, all peak lags are 
exceeded by 52--56 of 200 simulated lags, and all lag lower limits 
are exceeded by 105--106 of 200 simulated lags, so the null hypothesis 
cannot be rejected at high confidence. We conclude that the 
observation of changes in the softness ratio leading changes in 
the summed light curve by $\sim$5 ksec is tentative at best.

\subsection{Cross-spectral Properties}   %%%% sect 3.5

In addition to using the CCF, interband correlations can be 
characterized using the cross-spectrum; analogous to the fact 
that the PSD is the Fourier transform of the ACF, the cross-spectrum 
is the Fourier transform of the CCF. The cross-spectrum is a 
complex number. Its squared magnitude is used to define the 
coherence, which is defined and discussed in $\S$3.5.1. The 
argument of the cross-spectrum is the phase spectrum, which is 
defined and discussed further in $\S$3.5.2. Further details of 
the cross-spectrum can be found in e.g., VFN03 and Papadakis, 
Nandra \& Kazanas (2001).

\subsubsection{Coherence Function}       %%%%%% sect 3.5.1

The coherence function $\gamma^2$$(f)$ gives the fraction of 
mean-squared variability of one time series that can be attributed 
to the other, at a given temporal frequency. If the two time 
series are related by a simple, linear transfer function, then 
they will have unity coherence. The coherence is defined as the 
magnitude squared of the cross spectrum normalized by the product 
of each light curve's PSD. Here, we use the discrete versions 
of the cross-spectrum and coherence functions, given in 
$\S$5.2.2 of VFN03. To reduce scatter, the coherence is calculated 
by averaging the real and imaginary components of the cross-spectrum 
and the auto-spectra over $m$ consecutive frequencies, assuming that
the estimates at each temporal frequency are independent.
We correct the coherence function for the influence of Poisson noise
by following Vaughan \& Nowak (1997). Their equation 8 gives
the noise-corrected coherence and its uncertainty. However, it is
applicable only in the condition of high coherence and high powers; 
more specifically, the intrinsic variability power must be at least a 
factor of a few times $\sqrt{m}$ $\times$ $P_{{\rm Psn}}$.
For this condition to be met, it was necessary to 
rebin the short-term light curves to 5000 sec. 
For the two short-term light curves, the cross-spectrum and PSD
estimated were calculated for each revolution separately,
then combined and sorted by frequency before binning.

For the short-term data, the coherence function was calculated for the
soft--medium, soft--hard, and medium--hard bands, as well as for 
H1--H2 and H2--H3. For the medium- and long-term data, the 
coherence function was measured for all combinations of the 
H1, H2, H3 and H4 light curves. The long-term coherence functions 
represent the longest time scale coherence functions measured for 
an AGN to date. The temporal frequencies sampled span an order of 
magnitude above and below the high- and low-frequency PSD breaks 
modeled by M03. Coherence was binned to reduce scatter for all 
three time scales. The value of $m$ was usually 15, but increased slightly to 
20 for coherence functions measured from pairs of light curves 
that included the (relatively noisy) H4 band. This yielded only
one coherence estimate for the medium-term data and two coherence 
estimates for the short-term data, but 7--9 estimates for the 
long-term data. 

The combined results, spanning the temporal frequency range 
$6 \times 10^{-8} - 1 \times 10^{-4}$ Hz, are shown in Figure~7.
We note that the uncertainties on the coherence points,
calculated using the formulation of Vaughan \& Nowak (1997),
are in general agreement with the scatter of the unbinned 
coherence estimates (not plotted); scatter is generally larger 
for relatively lower coherence estimates. This suggests
that the coherence estimates are reasonably accurate. Overall, 
on the temporal frequencies sampled, the measured coherence 
of the light curves is relatively close to unity.
However, the coherence has a tendency to decrease somewhat as 
the relative separation of the two energy bands increases. 
This appears consistent with what was observed in the CCFs 
($\S$3.4). However, Poisson noise could play a role, artificially 
reducing the measured coherence.

As a preliminary check on the accuracy of the short time scale 
coherence measurements, coherence functions were also calculated 
using the MOS2 data. At the lowest frequency bin, the coherence 
functions derived from the MOS2 data agree with those derived 
from the pn data. However, the high level of Poisson noise in 
the MOS2 data prevents further comparison at higher temporal 
frequencies. As another check on the accuracy of the short 
time scale coherence measurements, the coherence was calculated 
between the MOS2 and the pn for each of the soft, medium, and hard 
bands. Because the light curves extracted from either detector 
should be identical except for the effects of Poisson noise, any 
deviation from unity coherence would have to be due to Poisson noise.
At frequencies lower than about 10$^{-4}$ Hz, the coherence is 
indeed close to 1 (not plotted).

Monte Carlo simulations were used as an additional check on the 
accuracy of the coherence estimates and to determine if any deviation 
from unity coherence can be attributed to the effect of Poisson noise.
Simulations were performed for each time scale separately. Two light 
curves were simulated using the same random number seed and identical PSD
shapes, to produce two light curves with intrinsic unity coherence. The 
best-fit medium-band high-frequency PSD slope and amplitude was assumed.
The simulated light curves for the two bands were resampled to match 
the observed data, and rescaled to the mean count rates in the softer 
and harder bands. Poisson noise was then added to the data according 
to the Poisson distribution (that is, randomly deviating the data using 
a Gaussian convolved with the square root of the observed mean error 
squared). One hundred sets of simulated data were produced. For each set, 
the coherence and its uncertainty was calculated as for the real data.
At most temporal frequencies (e.g., particularly the long- and 
medium-term data), the coherence was reasonably close to unity, and 
the uncertainty in the coherence was in reasonable agreement
with the scatter, indicating that the coherence estimation was 
reasonably accurate. The 90$\%$ and 95$\%$ lower limits to the coherence 
were determined, and are plotted in Figure~7. Any measured coherence 
point lying below these lines thus indicates a drop in coherence 
that is intrinsic to the source, and not an artifact of the Poisson 
noise, at 90$\%$ or 95$\%$ confidence. For most temporal frequencies, 
the artificial drop in coherence is small. At most temporal frequencies, 
particularly below $\sim$10$^{-6}$ Hz, the deviation of the measured 
coherence from unity is greater than the artificial drop due to Poisson 
noise at greater than 95$\%$ confidence. Furthermore, there is a 
general tendency for the intrinsic coherence to decrease as the 
energy separation of the bands increases.

%%% c     Deviate randomly using gau convolved with sqrt (observed mean error
%%% c     sqrd). This is VFN03 method and the method I use.
%%% c     McH04 does simming a l.c., deviating using gau*(sqrt(simmed mean error sqrd).

Previous analysis ($\S$3.3) has shown that the high-frequency PSD 
slope in NGC~3783 is consistent with being energy-dependent; such a 
dependence could contribute to a reduction in measured coherence. However, 
assuming that the PSD break frequencies change, and assuming that the 
power-law slopes below the high-frequency break do not change, 
any putative change in high-frequency slope will not have any 
effect on the coherence below $4 \times 10^{-6}$ Hz in NGC~3783. 
The Monte Carlo simulations were repeated exactly as above for the 
short-term data, but using the best-fit PSD slopes and 
amplitudes from $\S$3.3.1, and for the medium-term data, assuming 
power law slopes of --2.3, --2.2, --2.1, and --2.0 for the H1, H2, 
H3 and H4 bands, respectively. The effects of Poisson noise were 
again added to the simulated data. The drop in coherence due to having 
different high-frequency PSD slopes was minimal for all bandpass 
combinations, and virtually insignificant compared to the drop in 
coherence due to Poisson noise (not plotted).

We can conclude that the coherence is reasonably close to unity 
over the temporal frequencies sampled, but there are deviations 
from unity coherence present that are greater than those 
introduced by Poisson noise or by having different high-frequency 
PSD slopes. The fact that the coherence tends to drop as the 
energy separation of the bandpasses increases is consistent 
with the CCF results from $\S$3.4.

\subsubsection{Phase Lags}     %%%%sect 3.5.2

The phase spectrum indicates the average phase shift between 
the components of the two light curves, as a function of 
temporal frequency. Phase lag analysis is complementary to 
CCF analysis %%% (e.g., the CCF of the SR versus the summed 
band light curves), as the two methods are sensitive to the 
variability on different time scales. CCFs tend to be most 
sensitive to the largest amplitude variations in a time series, 
which, for red noise processes, will be those on relatively 
larger time scales (10 ksec and longer in the case of the \xmm\ 
light curves studied here). Phase lag analysis, meanwhile, is 
more sensitive to the variations on relatively smaller time scales 
after accounting for binning as described below (hundreds of seconds
in this case). However, sensible results for phase lag measurement 
requires data of sufficiently high quality. For instance, one 
requires continuous or near-continuous sampling to recover
phase lags that are a small fraction of the total duration.
The medium- and long-term sampling light curves, for instance,
each span only $\sim$2 dex in temporal frequency, and
the sparse sampling at the lowest temporal frequencies is
insufficient to recover e.g., phase lags that are a small 
fraction (a few percent) of the total duration. Consequently, 
here we concentrate on recovering phase lags from the short-term 
data only. Additionally, the coherence must be high (the variations 
in the two bands must be reasonably well-correlated) for lags to 
be meaningful. In the case of NGC~3783, phase lags are only 
recoverable in the narrow temporal frequency range of 
$6 \times 10^{-5} - 1.3 \times 10^{-4}$ Hz.

We follow section 5.3.1 of VFN03 to estimate the phase lags from 
the short-term light curves for NGC~3783. We combined unbinned data 
from revolutions 371 and 372 and sorted by frequency. Phase lag estimates 
were binned over 20 consecutive temporal frequencies to reduce scatter.
The phase lags $\phi(f)$, along with their uncertainties $\Delta\phi$, 
which were determined from equation 16 of Nowak \et\ (1999a), 
are plotted in Figure~8. A positive lag is defined as relatively 
soft band variations preceding those in the hard band. There is 
approximate closure in the resulting phase lags: the sum of the S--M 
and M--H phase lags at a given frequency bin is roughly equal to 
the S--H phase lags. There are only two frequency bins, but the data 
are not inconsistent with phase lags increasing as the energy separation 
of the bands increases and as temporal frequency increases, 
qualitatively consistent with the previous phase lag studies
of AGNs and XRBs. The dependence on temporal frequency explains 
the asymmetry present in the CCFs; this phenomenon has been noted 
previously by e.g., Maccarone \et\ (2000) and M$^{\rm c}$Hardy \et\ (2004).
The error on the phase lags was seen to be in reasonable agreement 
with the scatter of the unbinned phase lag estimates at these 
frequencies, indicating that the formulations used here for $\phi$ 
and $\Delta\phi$ are reasonable. The phase lags are consistent with 
being described by the relation $\phi(f) = C f^{-1.2}$, where 
$C$ $\sim$ 0.001, 0.003, and 0.004 for the S--M, M--H, and S--H 
lags, respectively, although these relations are obviously 
poorly constrained due to the narrow range of 
temporal frequencies sampled.             %%%%XXX plot unbinned phase lags,too?
The relation between phase lags and temporal frequency has 
been quantified in previous studies (see references above) 
as $\phi(f) = C f^{-1}$; in this case, the best-fit 
$f^{-1}$ relation yields $C$ $\sim$ 0.009, 0.026, and 0.036 
for the S--M, M--H, and S--H lags, respectively.

We follow section 4.2 of Nowak \et\ (1999a) to estimate the 
effective noise limit; their equation 14 yields the lower limit 
on the uncertainty on the phase lags due to the presence of
Poisson noise. These limits as a function of frequency were 
calculated for each bandpass pair using the average of the 
best-fit PSD shapes from $\S$3.3.1  (e.g., assume $\beta$ = 2.9 
for S--M, etc.), and the average of the power due to Poisson noise
from both bands and both revolutions of data. The measured coherence 
is consistent with, and assumed to have the form of 
$\gamma^2$$(f) = exp(-f/10^{-3.8} Hz)^2$; the cutoff frequency
was chosen somewhat arbitrarily and given the results of 
$\S$3.5.1.  These limits are shown in Figure 8, and indicate that 
phase lags cannot be recovered from the data at
temporal frequencies above $\sim$10$^{-3.8}$ Hz.

Phase lags were also recovered for the H1--H2, H2--H3, and 
H1--H3 bandpass pairs, but due to an even narrower usable 
temporal frequency range, this analysis yielded only one 
binned phase lag point for each pair. The phase lags obtained, 
at a binned frequency of 4.33 $\times$10$^{-5}$ Hz,
were reasonably consistent with closure: 
$\phi(f)$ = 300$\pm$10 s for the H1--H2 bands,
$\phi(f)$ = 500$\pm$30 s for the H2--H3 bands, and
$\phi(f)$ = 800$\pm$50 s for the H1--H3 bands.
The detection limits due to Poisson noise were in the 
vicinity of 10, 30, and 50 seconds for the H1--H2, 
H2--H3, and H1--H3 bands, respectively. The MOS2 data were 
tested for consistency, but found to be too noisy
for phase lags to be recovered.

Monte Carlo simulations, similar to those performed for the 
coherence functions above, were performed as an additional 
verification of the phase lag estimates. One hundred pairs of light 
curves were simulated, with each pair simulated with
the same random number seed and same underlying PSD shape
to ensure that they have unity coherence, identical Fourier phases,
and no intrinsic lag. Poisson noise was added to the simulated 
light curves as described in $\S$3.5.2. Phase lags were measured 
in a manner identical to the real data. The resulting simulated 
phase lags were reasonably close to zero, indicating that the observed 
phase lags are not an artifact due to Poisson noise. Simulations were 
also performed with light curves derived from different 
underlying PSD slopes. The resulting simulated phase lags were again close to zero,
suggesting that the energy-dependent PSD shape did not bias the phase lag results.

%%%%  SIMS: VFN03 doesn't bother with adding constant phase offset

\section{Discussion and Conclusions}     %%%%% Sect 4

\subsection{Summary of observational results}      %%% -------4.1
 
We have presented several probes of the energy-dependence of 
the variability properties of the Seyfert~1 galaxy NGC~3783,
using complementary {\it RXTE} monitoring and {\it XMM} long-looks 
to quantify the variability over a broad range of time scales. 
We find relatively softer energy bands to display higher amplitudes 
of variability on all time scales, in line with numerous previous 
results. On short time scales, we find that the high-frequency 
PSD slope is consistent with flattening as photon energy increases. 
Consequently, the central peak of the ACFs narrower as photon 
energy increases.  Cross-correlation functions show peaks near 
zero lag, but with a slight asymmetry in the sense that
the CCFs are skewed towards positive lags. The peak 
correlation coefficient decreases slightly as the energy 
separation of the bands increases. The hardness ratio is generally 
correlated with the continuum flux, consistent with spectral softening 
as flux increases, as is commonly seen in Seyferts. However, 
cross-correlation measurements of the SR versus the summed band 
tentatively reveal an indication for the softening to precede the 
brightening by very roughly 5 ksec. The broadband coherence 
function is close to unity in the frequency range 
$6 \times 10^{-8} - 1 \times 10^{-4}$ Hz; however, the 
coherence seems to decrease slightly as the energy 
separation of the bands increases. The temporal frequency 
dependence of the phase lags was explored on short time scales;
the data are consistent with phase lag increasing both as the 
temporal frequency decreases and as the energy separation of 
the two bands increases; this behavior explains the asymmetrical 
and skewed shape of the CCF.

\subsection {Comparison to previous results}   %%%%%%----- this is 4.2

All of these results are in qualitative agreement with previous 
studies of Seyferts and XRBs (See $\S$1 for objects and references), 
consistent with the notion that XRBs and Seyferts are mass- and/or luminosity- scaled 
versions of the same accretion phenomenon. However, a detailed comparison 
between NGC 3783 and XRBs would ideally take into consideration 
whether NGC 3783 is an analog of an XRB in its low/hard or high/soft state. 
In terms of broadband PSD shape, NGC~3783 is consistent with the low/hard state
of the Galactic black hole candidate Cyg X-1, since both PSDs are not 
dominated by 1/$f$ noise at the lowest temporal frequencies 
probed (down to $\sim$10$^{-8}$ Hz in the case of NGC~3783; 
also see Axelsson \et\ 2005). However, given NGC~3783's black 
hole mass of 3.0 $\pm$ 0.5 $\times$ 10$^{7}$ $\Msun$ (Peterson \et\ 2004),
the high-frequency PSD break of 4 $\times 10^{-6}$ Hz for 
NGC~3783 is more consistent with linearly scaling with black hole 
mass from Cyg X-1's high/soft state than from the low/hard state,
assuming that both objects accrete at the same rate relative 
to Eddington. The transition between the low/hard and high/soft states 
in XRBs usually occurs when the accretion rate is $\sim$2$\%$ 
of Eddington (Maccarone 2003); however, estimating the 
accretion rate of NGC~3783 does not shed any light on 
state identification. Given NGC~3783's 2--12 keV luminosity of 
$\sim 2 \times 10^{43}$ erg~s$^{-1}$ (Markowitz \& Edelson 2004),
the relation between X-ray luminosity and bolometric luminosity 
$L_{{\rm Bol}}$ given by Padovani \& Rafanelli (1988) of 
$L_{{\rm Bol}}$ = 60 $\times$ $\nu$$L_{\nu}$ at $\nu$=2 keV,
and a photon index of 1.7 (e.g., Markowitz, Edelson \& Vaughan 2003a), 
the bolometric luminosity is estimated to be $\sim 5 \times 10^{44}$ erg~s$^{-1}$
(Note that the factor 27 given as the ratio of $L_{{\rm Bol}}$ to 
2--12 keV luminosity in Markowitz \& Edelson 2004 is only valid 
if the photon index is close to 1.6). Given NGC~3783's black hole 
mass of 3 $\times$ 10$^{7}$ $\Msun$, this yields an estimate of 
$L_{{\rm Bol}}$/$L_{\rm Edd}$ of roughly 12$\%$. While this value 
is above the transition threshold in XRBs and implies a high/soft 
state analogue, it is not immediately reconcilable with PSD shape 
if the specific analogy to Cyg X-1 is valid. Furthermore, another 
twist is pointed out by Done \& Gierli$\grave{\rm n}$ski (2005): 
the PSDs of transient accreting compact sources (i.e., in contrast 
to the relatively more stable Cyg X-1) often are not dominated by 
1/$f$ noise at their lowest frequencies in either their low/hard or
high/soft state. Instead of labeling NGC~3783 as a low/hard or 
high/soft state accretor, it is therefore more conservative to simply 
say that this paper presents the first measurement of cross-spectral 
properties at low temporal frequencies for a Seyfert galaxy whose 
broadband PSD shows evidence for two breaks and is not dominated by 
1/$f$ noise at the lowest temporal frequencies probed.

%%%  Assuming linear scaling with black hole mass from the low/hard state of Cyg X-1, 
%%%  the temporal frequencies sampled for NGC~3783 are equivalent to  0.2 -- 300 Hz in Cyg X-1.

NGC~3783's coherence is seen to be close to unity in the vicinity 
of both the high-frequency and low-frequency PSD breaks, as well 
as in between these breaks. This is similar behavior to that of the 
coherence function for Cyg X-1 in the low/hard state (Nowak \et\ 1999a).
The results so far for objects with both measured coherence 
and PSD breaks seem to indicate that coherence is reasonably 
close to unity below the high-frequency break in most objects. 
This includes Cyg X-1 and GX 339--4 in the low-hard state 
(Nowak \et\ 1999a, 1999b), as well as MCG--6-30-15 and NGC~4051 
(VFN03 and M$^{\rm c}$Hardy \et\ 2004, respectively), though 
it has been noted that the PSD shapes of these latter two 
objects are more similar to Cyg X-1 in the high/soft state. The 
narrow-line Seyfert 1 galaxy Mkn~766 may be a possible exception 
in this regard, as discussed by Vaughan \& Fabian (2003).

Further similarities between NGC~3783 and the low/hard state of 
Cyg X-1 can be found in the ACFs and CCFs. Maccarone \et\ (2000) 
noted that the FWHMs of the central ACF peaks scaled as $E^{-0.2}$ 
($E$ is the photon energy) in the low/hard state of Cyg X-1. The 
ACF FWHMs for NGC~3783 are roughly 42 ksec, 36 ksec and 26 ksec, 
respectively, for the S, M and H bands in the revolution 371 data,
and roughly 48 ksec, 40 ksec and 34 ksec, respectively, for S, M, 
and H in revolution 372, and these FWHMs are also consistent with 
scaling roughly as $E^{-0.2} - E^{-0.3}$. Additionally, Maccarone \et\ 
(2000) noted that the CCF peak lags in Cyg X-1 were less than 2 msec. 
If we scale by a factor of $3 \times 10^6$, the ratio of the black hole 
masses, one would expect peak CCF lags in NGC~3783 to be within 
roughly 6 ksec, as observed. However, we note that the CCF (and 
the ACF) can change character (e.g., FWHMs) quite dramatically 
from one light curve realization of a given PSD to the next, so 
a CCF comparison should be regarded with caution. Nonetheless, the 
data are consistent with the notion of similar variability 
properties in NGC~3783 and (at least the low/hard state of) Cyg X-1.

\subsection{Physical implications of coherence results}   %%%% this is section 4.3 

As explained in Vaughan \& Nowak (1997), coherence between two 
energy bands can be lost if the transfer function is nonlinear, 
or if there are multiple, uncorrelated transfer functions. In the 
case of multiple flaring regions (e.g., shot-noise models), if 
more than one region contributes to the signal in both energy bands, 
then it is possible for coherence to be low, even if individual regions 
produce perfectly coherent variability. This is because in order to 
achieve unity coherence, each separate region independently requires 
the same linear transfer function that links soft to hard variations. 
The relevance of coherence to shot-noise models is discussed further 
in $\S$4.4 below.

%%%%Unity or high coherence thus requires linear transfer functions, and usually requires
%%%%localized sources and responses, a temporally uniform source, or a uniform response throughout the disk

%%%%More mechanisms exist for destroying coherence than preserving coherence.

The high level of coherence on temporal frequencies below the 
high-frequency PSD break in NGC~3783 may indicate that the corona 
is effectively static on these time scales. More specifically, 
the fact that coherence is high for temporal frequencies near the 
low-frequency PSD break may indicate that the time scale 
corresponding to the low frequency PSD break is not directly 
associated with the dynamic time scale of the corona.

Drops in coherence have been seen at temporal frequencies higher 
than the high-frequency PSD break in Cyg X-1, MCG--6-30-15, and NGC~4051
(Nowak \et\ 1999a, VFN03, M$^{\rm c}$Hardy \et\ 2004). The corona may 
be dynamic on those short time scales; that is, the loss of 
coherence could indicate that these time scales may be directly 
associated with the formation time scales and/or changes in 
temperature and/or physical structure of the corona (e.g., 
Nowak \& Vaughan 1996, Nowak \et\ 1999a). No such coherence drop 
is detected in the present data for NGC~3783 due to insufficient 
variability-to-noise. The highest temporal frequency bin in the 
short-term data (at $1 \times 10^{-4}$ Hz) may, at first glance, be 
suggestive of the beginning of a drop in intrinsic coherence, but 
the noise-corrected coherence and the simulations 
show that this drop is instead likely an artifact due to Poisson noise.

\subsection{Phenomenological variability models}   %%%%% Section 4.4

%%%%%%%%%%%%%%%%%%%%%%%%%%%%%%%%%%%%%%%%%%%%%%%%%%%%%%%%%%%%%%%%%%%%%%%%%%%%%%%%%%%%%%%%%%%

Many models have been invoked to explain the red noise variability 
properties of Seyferts and XRBs, including shot-noise models (e.g., 
Merloni \& Fabian 2001), rotating hot-spots on the surface of the
accretion disk (Bao \& Abramowicz 1996), self-organized criticality 
(``pulse avalanche'' models; Mineshige, Ouchi \& Nishimori 1994), 
magnetohydrodynamical instabilities in the disk (Hawley \& Krolik 2001), 
and inwardly-propagating fluctuations in the local accretion rate 
(Lyubarskii 1997). As noted by e.g., M$^{\rm c}$Hardy \et\ 2004, 
shot-noise models can reproduce virtually any red-noise PSD shape. However, on short time scales,
shot-noise models have difficulty reproducing the linear relation 
between rms variability and continuum flux observed in XRBs and Seyferts 
(Uttley \& M$^{\rm c}$Hardy 2001, Uttley, M$^{\rm c}$Hardy \& Vaughan 
2005). Furthermore, as discussed in Vaughan \& Nowak (1997) and $\S$4.3, 
shot-noise models lead to low observed coherence unless each separate 
emitting region independently has the same linear transfer function 
between soft and hard energy bands. For example, in the ``thundercloud'' 
model of Poutanen \& Fabian 1999, phase lags could be attributed to the 
spectral hardening of individual flares (see also B\"{o}ttcher \& Liang 
1998), but unity coherence can be achieved only if the exact same 
Comptonization spectrum and flare evolution independently apply to 
each flare (Nowak \et\ 1999b).

Thermal Comptonization of soft seed photons by a hot corona is a 
likely explanation for the X-ray emission. It is supported e.g., 
by X-ray energy spectra, as well as by the correlation between 
UV flux and X-ray photon index observed in NGC~7469 by Nandra \et\ 
(2000). The phase lags could thus be attributed to the time scale 
for the seed photons to diffuse through the corona and undergo 
multiple up-scatterings, and, as discussed by e.g., VFN03, the 
simplest models predict that hard photons should lag soft photons. 
%%%   In the case of NGC~3783, the soft-to-hard X-ray lags observed are 
%%%   on the order of 100--1000 s. If these lags correspond to light-crossing 
%%%   time scales, then, given NGC~3783's black hole mass of 
%%%   $3 \times 10^{7} \Msun$ (Peterson \et\ 2004), they correspond to radii less 
%%%   than 6 $R_{\rm g}$ ($R_{\rm g}$ = G$\mbh$/c$^2$), e.g., consistent with 
%%%   an origin in the inner regions of a Comptonizing corona.
However, the simplest Compton models do not predict that the lags should 
depend on temporal frequency (Miyamoto \et\ 1998), as observed. Furthermore, 
if phase lags do scale with the inverse of the temporal frequency, then 
extension of this relation to low frequencies would require a non-compact 
($\sim$ 10$^{3-5}$ gravitational radii; see Nowak \et\ 1999b and Papadakis, 
Nandra \& Kazanas 2001) corona (e.g., Kazanas, Hua \& Titarchuk 1997). 
In addition, Compton scattering models predict that for relatively higher 
energy photons, which have undergone more scatterings, the most rapid variations 
will be washed out. This would lead to high-frequency PSD slopes that
steepen as photon energy increases, contrary to observations. 
However, the energy dependence of the PSD could be obtained with a corona whose
temperature increased towards smaller radii (Kazanas, Hua \& Titarchuk 1997).
%%%%  need to add anything to this last sentence?

%%%%%%%     Lyu 97  Kot 01   Chur 01  models

A quantitative model in which inwardly-propagating variations in the 
local mass accretion rate are responsible for the observed X-ray 
variability seems to be able to explain many of the observational 
results in Seyferts and XRBs (Lyubarskii 1997, Churazov, Gilfanov 
\& Revnivtsev 2001), Kotov \et\ 2001; see also discussions in VFN03 and 
M$^{\rm c}$Hardy \et\ 2004). In this model, variations at a 
given radius are associated with the local inward propagation time 
scale, so relatively smaller radii are associated with relatively 
more rapid variations. Churazov \et\ (2001) have suggested a geometrical 
configuration wherein an optically-thick, geometrically-thin soft disk is 
sandwiched above and below by an optically-thin corona. The disk is 
responsible for the bulk of the accretion rate but is not strongly variable; 
the bulk of the variations occur in the coronal flow. These variations 
propagate inward until they reach, and modify the emission of, the central 
X-ray emitting region, yielding a 1/$f$ X-ray PSD across a broad range of 
temporal frequencies (Kotov \et\ 2001, Lyubarskii 1997). The most rapid 
variations are suppressed, leading to a PSD break at high temporal frequencies 
(e.g., at the frequency $f_3$ in Figure 3 of Churazov \et\ 2001); the break 
time scale may possibly be associated with the viscous time scale at the 
edge of the central X-ray emitting region. In the high-soft state, the 
disk extends down to the central X-ray emitting region. In the low-hard state, 
the disk is truncated; below the truncation radius, all of the accretion flow is 
in the form of the corona. Local variations at all radii will contribute to a 
1/$f$ X-ray PSD. However, variations originating from radii larger than the 
disk truncation radius are suppressed due to the presence of the non-variable disk. 
The resulting X-ray PSD follows a 1/$f$ relation over two separate ranges of 
frequencies, but linked by a 'transition region' consisting of an f$^0$ 
dependence over some intermediate range of frequencies; variations at these 
temporal frequencies are not due to local instabilities in the accretion flow 
but are rather the result of stochastic superposition of variations on shorter 
time scales. This model thus incorporates two additional PSD breaks at low 
frequencies, whereby the PSD slope changes from --1 to 0, then back to --1
(temporal frequencies $f_1$ and $f_2$ in Figure 3 of Churazov \et\ 2001, with
$f_1 < f_2$). The "low-frequency break" at $\sim 2 \times 10^{-7}$ Hz in 
NGC 3783, for instance, could be the characteristic frequency in the coronal 
flow at the disk truncation radius. The lowest PSD break in this model ($f_1$),
below which the PSD returns to a 1/$f$ shape, has not yet been observed in 
NGC~3783 or any AGN; from scaling with Cyg X-1, it might be expected near 
$\sim$10$^{-9}$ Hz (corresponding to a time scale of $\sim$30 years) in NGC 3783.
Another common method of modeling XRB PSDs, particularly those in the low/hard
state, is with Lorentzian components; in the context of this class of models,
such components could be associated with the corona at radii below the disk 
truncation radius and thus are absent from the high frequency PSDs of XRBs 
in the high/soft state. The "two-break" PSD shape for NGC 3783 may actually 
consist of two Lorentzian components if NGC 3783 is a low/hard state AGN. 

The accretion rate variation model class is qualitatively consistent 
with the rms-flux relation seen in Seyferts and XRBs. In the case of NGC~3783, 
there is an insufficient flux range on both medium and long time scales 
to test for the rms-flux relation (the variability is dominated by 
variations below the high-frequency break), and there is not enough
data on short time scales after suitably binning the data. Kotov 
\et\ (2001) suggested that the spectrum can be a function of radius, 
with relatively harder emission emanating from smaller radii, and 
therefore associated with more rapid variability; this leads to 
high-frequency PSD slopes that flatten versus energy. Relatively 
longer temporal frequencies are associated with larger radii and 
longer characteristic time scales, hence phase lags are longer. 
Phase lags increase as the energy separation of the bands increases
because the two bands are associated with two radii that are 
relatively more spatially separated. CCF peak correlation 
coefficients and coherence both decrease as the energy separation of 
two bands increases because propagations can only travel inward, 
and the relatively harder energy band (emitted at a smaller radius) 
will have undergone additional perturbations. CCF lags are expected 
to be close to zero, as observed, since the bulk of the X-rays are 
emitted at very small radii, but the dependence of the phase lags
on temporal frequency causes the CCF asymmetry.

%%%This model also explains the drop in coherence above the break seen in several Seyferts (though 
%%%%the current \xmm\ data for NGC~3783 are not adequate to test for this drop):

%%%%Perturbations can only propagate inwards in this model. Lower energies associated with a certain radius
%%%do not see shorter period variations generated closer in to the BH. 
%%%highest energies see largest spectrum of variations

%%%%%%%%%%%%%%%%%  Kording & Falcke pivoting PL model

Finally, K\"{o}rding \& Falcke (2004) note that a pivoting power-law model 
seems qualitatively capable of reproducing many of the observed 
variability characteristics in XRBs (and by extension, Seyferts).  
This model yields high coherence over a broad range of time scales,
ACF peaks which get narrower as energy increases, and phase lags 
that exhibit a dependency on temporal frequency in a manner similar 
to those observed.

\acknowledgments 
A.M.\ thanks Shai Kaspi for providing both the {\it Chandra} light curves
and useful suggestions, and Phil Uttley for useful discussions.
This work has made use of observations obtained with {\it XMM-Newton}, an ESA science mission with instruments and contributions directly funded by ESA member states and the US (NASA). This work has made use of data obtained through the High Energy
Astrophysics Science Archive Research Center Online Service, provided by
the NASA Goddard Space Flight Center, and the NASA$/$IPAC Extragalactic Database which is
operated by the Jet Propulsion Laboratory, California Institute of
Technology, under contract with the National Aeronautics and Space
Administration.

%%%%%%%%%%%%%%%%%%%%%%%%%%%%%%%%%%%%%%%%%%%%%%%% TABLES %%%%%%%%%%%%%%%

\begin{deluxetable}{lcccccc}
\tabletypesize{\footnotesize}
\tablewidth{6.5in}
\tablenum{1}
\tablecaption{Source and derived variability parameters\label{tab1}}
\tablehead{
\colhead{Time}  & \colhead{}  &\colhead{Bandpass}& \colhead{Mean}        & \colhead{} & \colhead{$P_{{\rm Psn}}$} & \colhead{$\fv$}  \\
\colhead{Scale} &\colhead{$N$} &\colhead{(keV)}   & \colhead{ct s$^{-1}$} & \colhead{S/N}  & \colhead{(Hz$^{-1}$)}     & \colhead{($\%$)}  } 
\startdata
Short (Rev.\ 371)& 108         & 0.2--12         & 9.05  & 102.3 & 0.22 & 15.2 $\pm$ 0.1\\
                 &             & 0.2--0.7        & 2.57  &  54.7 & 0.80 & 20.4 $\pm$ 0.2\\
                 &             & 0.7--2          & 3.71  &  66.1 & 0.55 & 15.4 $\pm$ 0.1\\
                 &             & 2--12           & 2.78  &  57.0 & 0.73 & 10.9 $\pm$ 0.2\\
                 &             & 2--4            & 1.37  &  40.1 & 1.47 & 12.2 $\pm$ 0.2\\
                 &             & 4--7            & 1.09  &  36.1 & 1.85 & 9.9 $\pm$  0.3\\
                 &             & 7--12           & 0.41  &  25.1 & 4.96 & 5.5 $\pm$  0.4\\
Short (Rev.\ 372)& 103         & 0.2--12         & 12.42 & 121.4 & 0.16 & 12.9 $\pm$ 0.1\\
                 &             & 0.2--0.7        & 3.63  &  65.9 & 0.55 & 15.9 $\pm$ 0.2\\
                 &             & 0.7--2          & 5.11  &  78.4 & 0.39 & 13.0 $\pm$ 0.1\\
                 &             & 2--12           & 3.70  &  66.6 & 0.54 & 10.7 $\pm$ 0.1\\
                 &             & 2--4            & 1.86  &  47.1 & 1.08 & 11.7 $\pm$ 0.2\\
                 &             & 4--7            & 1.42  &  41.4 & 1.41 & 9.8 $\pm$  0.2\\
                 &             & 7--12           & 0.48  &  26.5 & 4.15 & 5.9 $\pm$  0.4\\
Medium           & 151         & 2--18           & 6.84 & 42.2 & 6.18   & 11.9$\pm$0.2\\       
                 &             & 2--12           & 6.33 & 42.5 & 6.18   & 12.1$\pm$0.2\\
                 &             & 2--4            & 1.61 & 22.8 & 23.4   & 13.9$\pm$0.4\\
                 &             & 4--7            & 2.77 & 29.1 & 13.1   &  11.9$\pm$0.3\\ 
                 &             & 7--12           & 1.95 & 22.4 & 23.0   & 11.0$\pm$0.4\\
                 &             & 12--18          & 0.51 & 8.4  & 166    & 9.4$\pm$1.5\\
Long             & 281         & 2--18           & 7.86 & 62.4 & 114    & 21.0$\pm$0.1\\
                 &             & 2--12           & 7.28 & 62.4 & 114    & 21.3$\pm$0.1\\
                 &             & 2--4            & 1.88 & 33.7 & 428    & 24.5$\pm$0.2\\
                 &             & 4--7            & 3.21 & 44.8 & 243    & 21.6$\pm$0.2\\
                 &             & 7--12           & 2.18 & 32.8 & 439    & 19.0$\pm$0.2\\
                 &             & 12--18          & 0.58 & 12.7 & 3010   & 18.7$\pm$0.6\\
\enddata
\tablecomments{Values are for the \xmm\ pn (short-term) 
and {\it RXTE} (medium- and long-term) data.
The \xmm\ pn data has been binned to 1200 s.             %%%%XTE-med binned to 22.9 ksec    XTE long binned to 4.3 d
$N$ (Col.\ [2]) is the number of points in each light curve.
For Col. (4), the entries in the medium- and long-term rows refer to {\it RXTE} mean count rate per PCU.
Col.\ (5) is the average signal-to-noise.
Col.\ (6) is the expected power due to Poisson noise. }
\end{deluxetable}

\begin{deluxetable}{lccc}
\tabletypesize{\footnotesize}
\tablewidth{6.5in}
\tablenum{2}
\tablecaption{Results for Monte Carlo PSD Fitting\label{tab2}}
\tablehead{
\colhead{Bandpass} & \colhead{Best-fit} & \colhead{Best-fit} & \colhead{Likelihood of} \\ 
\colhead{(keV)}    & \colhead{$\beta$}      & \colhead{$A$ (Hz $^{-1}$)}& \colhead{acceptance}}
\startdata
0.2--12   & 2.60$^{+1.40}_{-0.80}$   &  7.2$^{+2.6}_{-1.8} \times 10^{3}$  & 0.84 \\            
0.2--0.7  & 3.15$^{+0.85}_{-0.80}$   & 13.9$^{+6.1}_{-4.4} \times 10^{3}$  & 0.55 \\            
0.7--2    & 2.60$^{+1.40}_{-0.95}$   &  8.5$^{+3.0}_{-2.2} \times 10^{3}$  & 0.87 \\            
2--12     & 2.20$^{+0.80}_{-0.65}$   &  4.3$^{+1.2}_{-1.0} \times 10^{3}$  & 0.95 \\            
\enddata
\tablecomments{Results from fitting the PSDs with an unbroken power law model (i.e.,
a break at 4.0$\times$10$^{-6}$ Hz, below the minimum frequency sampled here). $A$ is the
amplitude at the 4.0$\times$10$^{-6}$ Hz break. Upper limits to $\beta$ are not constrained.}
\end{deluxetable}

\begin{deluxetable}{lccc}
\tabletypesize{\footnotesize}
\tablewidth{6.5in}
\tablenum{3}
\tablecaption{Results for $\chi^2$ PSD Fitting\label{tab3}}
\tablehead{
\colhead{Band} & \colhead{Best-fit} & \colhead{Best-fit} & \colhead{} \\ 
\colhead{(keV)}& \colhead{$\beta_{{\rm obs}}$} & \colhead{$A$ (Hz $^{-1}$)}& \colhead{$\chi^2$/d.o.f.}}
\startdata
0.2--12  & 2.43$\pm$0.05 &  12.9$^{+4.9}_{-3.6}$ $\times$ 10$^{3}$ & 4.0 / 16 \\
0.2--0.7 & 2.70$\pm$0.06 &  24.5$^{+11.8}_{-7.9}$ $\times$ 10$^{3}$ & 5.3 / 16 \\
0.7--2   & 2.40$\pm$0.05 &  13.2$^{+5.0}_{-3.7}$ $\times$ 10$^{3}$ & 3.7 / 16 \\
2--12    & 2.21$\pm$0.04 &  7.4$^{+2.4}_{-1.8}$ $\times$ 10$^{3}$ & 4.5 / 16 \\
\enddata
\tablecomments{Results of least-squares fitting to the observed \xmm\ PSDs. 
Note that these values do not correspond directly to the underlying PSD model shape
parameters due to the presence of red noise leak. Errors on $\beta_{{\rm obs}}$
correspond to a change in $\chi^2$ of $\Delta$$\chi^2$ = 2.71. }
\end{deluxetable}

\begin{deluxetable}{lccccccc}  
\tabletypesize{\footnotesize}
\tablewidth{6.5in}
\tablenum{4}
\tablecaption{Cross Correlation Function Results\label{tab4}}
\tablehead{
\colhead{Time} &\colhead{Band }  &\colhead{Band}& \colhead{ICF} & \colhead{ICF}  &\colhead{DCF}&\colhead{DCF}&\colhead{}\\
\colhead{Scale}&\colhead{1}&\colhead{2} &\colhead{$r_0$}&\colhead{$\tau$}&\colhead{$r_0$}&\colhead{$\tau$}& \colhead{$F_{{\rm sims}}$}}
\startdata
Short (Rev.\ 371) & S & M & 0.98 & +0.6$\pm$0.3 ks & 0.98 & 0 ks    &  59/200 \\
                  & S & H & 0.92 & +3.0$\pm$0.8 ks & 0.90 & +2.4 ks &  17/200 \\
                  & M & H & 0.95 & +1.8$\pm$0.6 ks & 0.94 & +1.2 ks &   0/200 \\
Short (Rev.\ 372) & S & M & 0.97 & +0.6$\pm$0.5 ks & 0.96 & 0 ks    &  11/200   \\
                  & S & H & 0.88 & +0.6$\pm$0.5 ks & 0.87 & 0 ks    &   6/200   \\
                  & M & H & 0.96 & +0.6$\pm$0.4 ks & 0.96 & 0 ks    &   1/200   \\
Medium            & H1  & H2 & 0.95 & 0$\pm$0.03 d   & 0.95 & 0 d         & 9/200    \\ %%%%% 22.9 ksec binning
                  & H1  & H3 & 0.90 & 0$\pm$0.03 d   & 0.90 & 0 d         & 0/200   \\
                  & H1  & H4 & 0.70 & 0$\pm$1.7  d   & 0.70 & 0 d         & 33/200   \\
                  & H2  & H3 & 0.90 & 0$\pm$0.05 d   & 0.90 & 0 d         & 0/200   \\
                  & H2  & H4 & 0.70 & 0$\pm$2.2  d   & 0.70 & 0 d         & 26/200   \\
                  & H3  & H4 & 0.73 & 0$\pm$1.9  d   & 0.73 & 0 d         & 77/200   \\
Long              & H1  & H2 & 0.93 & 0$\pm$0 d & 0.93 & 0 d              & 0/200    \\ 
                  & H1  & H3 & 0.94 & 0$\pm$0 d & 0.94 & 0 d              & 0/200   \\ 
                  & H1  & H4 & 0.80 & 0$\pm$0 d & 0.80 & 0 d              & 0/200   \\ 
                  & H2  & H3 & 0.95 & 0$\pm$0 d & 0.95 & 0 d              & 0/200   \\ 
                  & H2  & H4 & 0.83 & 0$\pm$0 d & 0.83 & 0 d              & 0/200   \\ 
                  & H3  & H4 & 0.84 & 0$\pm$0 d & 0.84 & 0 d              & 0/200   \\ 
\enddata
\tablecomments{Lags are defined such that a positive lag means band 2, the harder band, is delayed with respect to
band 1, the softer band. The DCF bin size was the resolution of the light curve, e.g., 1200 s for the short-term data; the
ICF resolution was one-half the DCF bin size. $F_{{\rm sims}}$ (Col.\ [8]) is 
the fraction of simulated light curves with $r_0$ greater than the observed value of $r_0$;
this fraction is the confidence level at which the null hypothesis of the two light curves
having unity coherence and a peak correlation 
coefficient of 1, with any lack of perfect correlation being
due solely to Poisson noise, can be rejected.}
\end{deluxetable}

\begin{deluxetable}{lcccccccc}  
\tabletypesize{\footnotesize}
\tablewidth{6.5in}
\tablenum{5}
\tablecaption{Cross Correlation Function Results, SR vs.\ Summed light curves \label{tab5}}
\tablehead{
\colhead{Time} &\colhead{Light}  &\colhead{Light}& \colhead{ICF} & \colhead{ICF}  &\colhead{DCF}&\colhead{DCF}&\colhead{$F_{{\rm sims}}$}&\colhead{$F_{{\rm sims}}$}   \\
\colhead{Scale}&\colhead{Curve 1}&\colhead{Curve 2} &\colhead{$r_0$}&\colhead{$\tau$}&\colhead{$r_0$}&\colhead{$\tau$}& \colhead{(peak lag)} & \colhead{(lag lower limit)}   }
\startdata
Short (Rev.\ 371) & S/M & S+M & 0.89 & +5.4$\pm$1.7 ks & 0.87 & 0 ks    &   54/200 & 63/200\\
                  & S/H & S+H & 0.90 & +6.6$\pm$1.9 ks & 0.89 & +2.4 ks &   18/200 & 40/200\\
                  & M/H & M+H & 0.83 & +7.8$\pm$4.1 ks & 0.82 & +1.2 ks &    4/200 & 18/200 \\

Short (Rev.\ 372) & S/M & S+M & 0.69 & +7.8$\pm$8.5 ks & 0.69 & +7.2 ks    &   52/200 & 105/200  \\
                  & S/H & S+H & 0.71 & +5.4$\pm$6.0 ks & 0.71 & +6.0 ks    &   56/200 & 106/200  \\
                  & M/H & M+H & 0.66 & +4.2$\pm$5.0 ks & 0.65 & +3.6 ks    &   55/200 & 105/200   \\

Medium            & H1/H2  & H1+H2 & 0.48 & --0.13$\pm$ 4.9 d   & 0.44 & 0 d         &  &    \\ %%%%% 22.9 ksec binning
                  & H1/H3  & H1+H3 & 0.55 & --0.13$\pm$ 4.9 d   & 0.49 & 0 d         &  &  \\
                  & H1/H4  & H1+H4 & 0.48 &  +0.13$\pm$ 5.3 d   & 0.40 & +0.265 d    &  &    \\
                  & H2/H3  & H2+H3 & 0.27 & --0.13$\pm$ 5.8 d   & 0.23 & 0 d         &  &  \\
                  & H2/H4  & H2+H4 & 0.33 &  +0.13$\pm$ 6.1 d   & 0.28 & +0.265 d    &     &    \\
                  & H3/H4  & H3+H4 & 0.32 &  +0.13$\pm$ 6.1 d   & 0.31 & +0.265 d    &     &    \\

Long              & H1/H2  & H1+H2 & 0.36 &    0$\pm$260 d & 0.36 & 0 d              & &    \\ 
                  & H1/H3  & H1+H3 & 0.62 &    0$\pm$250 d & 0.62 & 0 d              & &   \\ 
                  & H1/H4  & H1+H4 & 0.45 &    0$\pm$150 d & 0.45 & 0 d              & &   \\ 
                  & H2/H3  & H2+H3 & 0.41 &    0$\pm$270 d & 0.41 & 0 d              & &   \\ 
                  & H2/H4  & H2+H4 & 0.30 & +2.1$\pm$400 d & 0.29 & 0 d              & &   \\ 
                  & H3/H4  & H3+H4 & 0.11 & +221$\pm$370 d & 0.13 & --123 d          &    &    \\ 
\enddata
\tablecomments{Lags are defined such that a positive lag means light curve 2, the summed light curve, is delayed with respect to
light curve 1, the SR light curve. The DCF bin size was the resolution of the light curve, e.g., 1200 s for the short-term data; the
ICF resolution was one-half the DCF bin size. $F_{{\rm sims}}$ (Col.\ [8] and [9]) show 
the fraction of simulated CCFs with a lag greater than the ICF peak lag and lag lower limit, respectively.}
\end{deluxetable}

%%% %%%%%%%%%%%%%%%%%%%%%%%%%%%  FIGURES %%%%%%%%%%%%%%%%%%%%%%%%%%

\begin{figure}                  %%%%% Figure 1 = total band light curves
\epsscale{0.99}
\plotone{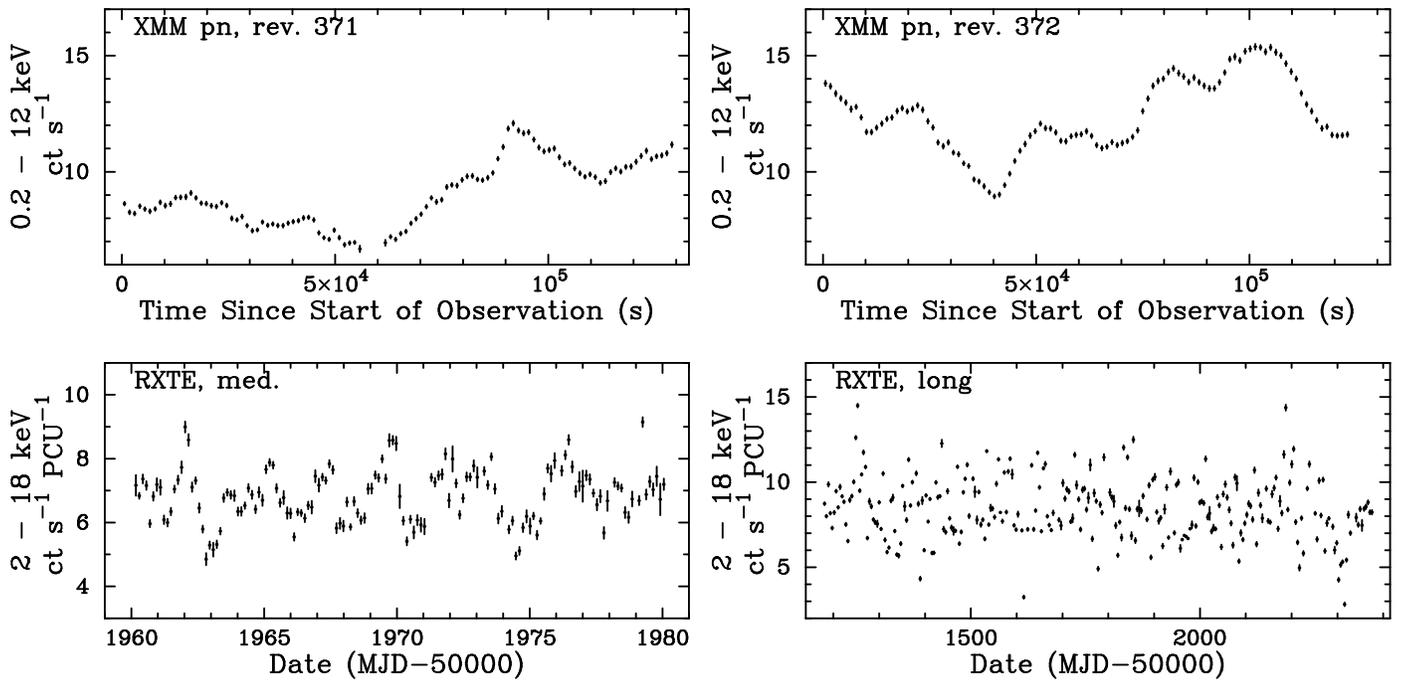}
\caption{Total bandpass light curves for all time scales.
Upper left: the revolution 371 \xmm\ pn 0.2--12 keV light curve, which started on 2001 December 17 at 19:41 UTC. 
Upper right: the revolution 372 \xmm\ pn 0.2--12 keV light curve, which started on 2001 December 19 at 19:33 UTC. 
Both pn light curves are binned to 1200 s. Note that the y-axes for both pn panels have the same scale.
Lower left: 2--18 keV \xte\ medium-term sampling light curve. 
Lower right: 2--18 keV \xte\ long-term sampling light curve.}
\end{figure}

%%%% New figure 2 is the SR light curve overplots!!!

\begin{figure}
\epsscale{0.99}
\plotone{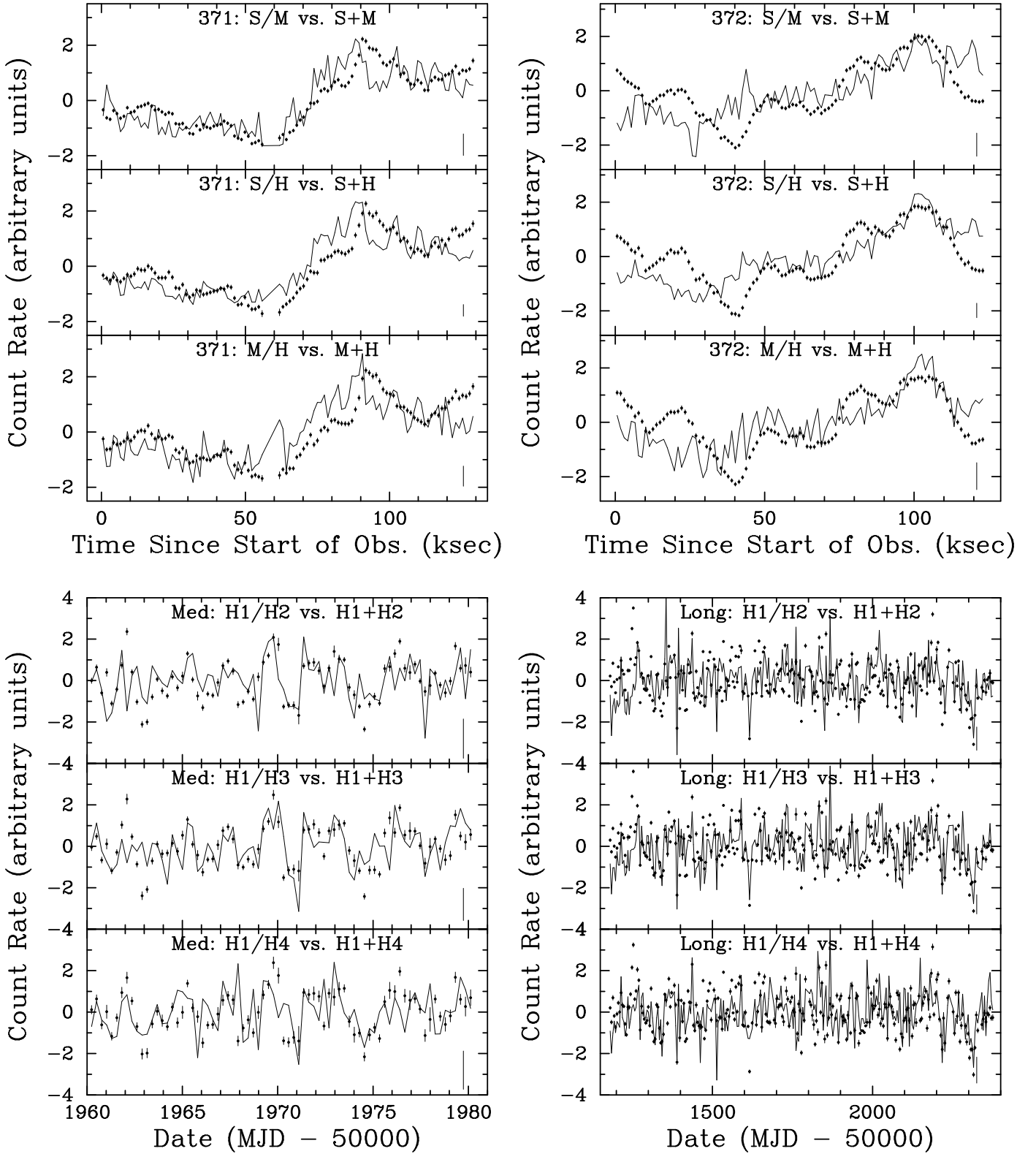}
\caption{Softness ratio (SR) and summed band overplots. 
The solid lines denote SR light curves;
the circles denote the summed band light curves. All light curves
have been mean-subtracted and normalized by their standard deviation. 
For clarity, errors on the SR light curves are not plotted; 
the error bar in the lower right corner of each panel denotes the
average two-sided error. There is a 
tentative tendency for the source to soften before it brightens on time scales
of $\sim$ 5ksec.}
\end{figure}

\begin{figure}            %%%% Figure 3 = MC PSD results           plots/fig2subt.eps
\epsscale{0.99}
\plotone{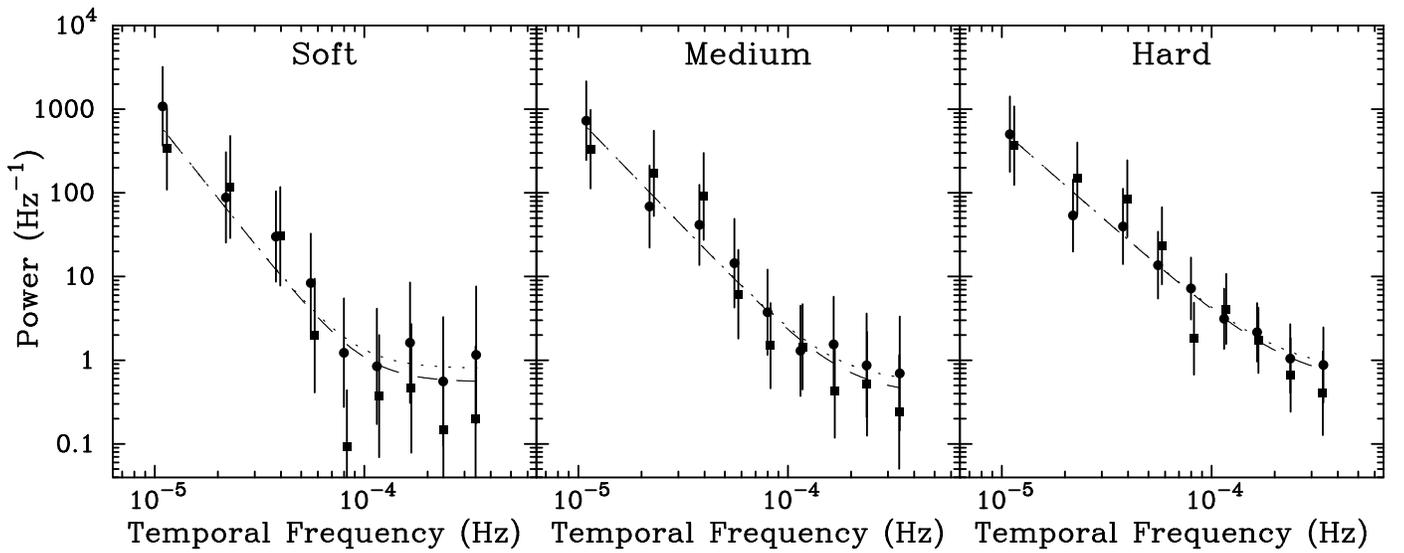}
\caption{The best-fit PSD model shape for the 
soft, medium, and hard-bands are shown for rev.\ 371 (dotted line) and rev.\ 372 (dashed line). The effects of
red-noise leak have been subtracted off.  
The symbols (circles for rev.\ 371 and squares for rev.\ 372) represent the difference between the
average distorted model PSD and the observed PSD, plotted with respect to the
underlying PSD model shape. The data are consistent with relatively softer bands having intrinsically steeper PSD slopes.}
\end{figure}

\begin{figure}        %%%%  Figure 4 = CCF results
\epsscale{0.99}
\plotone{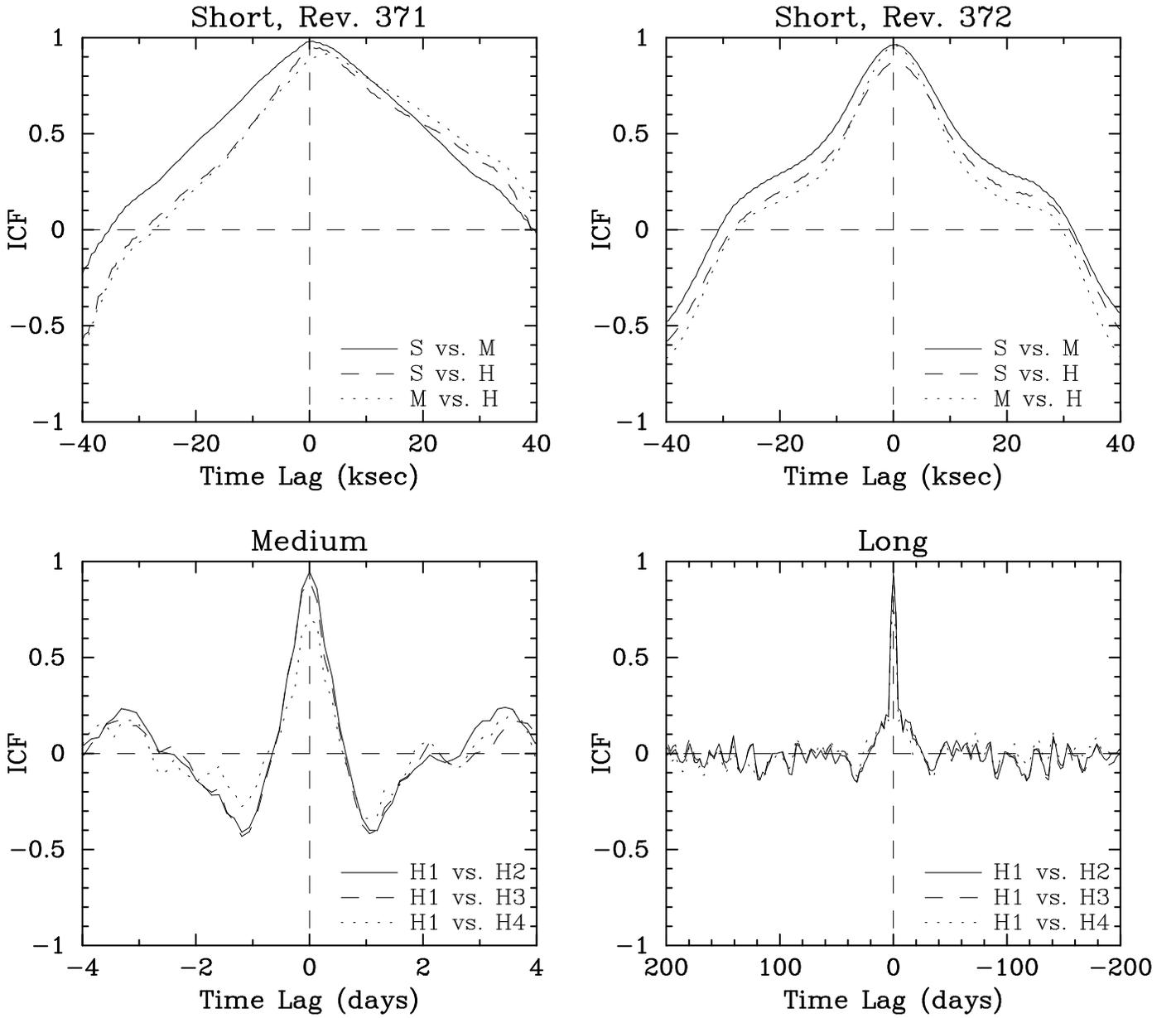}
\caption{Cross-correlation functions are shown for all time scales. A positive lag indicates that hard photon
variations lag those in the soft band. For clarity, the 
H2-H3, H3-H4 and H2-H4 medium- and long-term CCFs are not plotted. }
\end{figure}

\begin{figure}          %%%% Figure 5 = ACF results
\epsscale{0.99}
\plotone{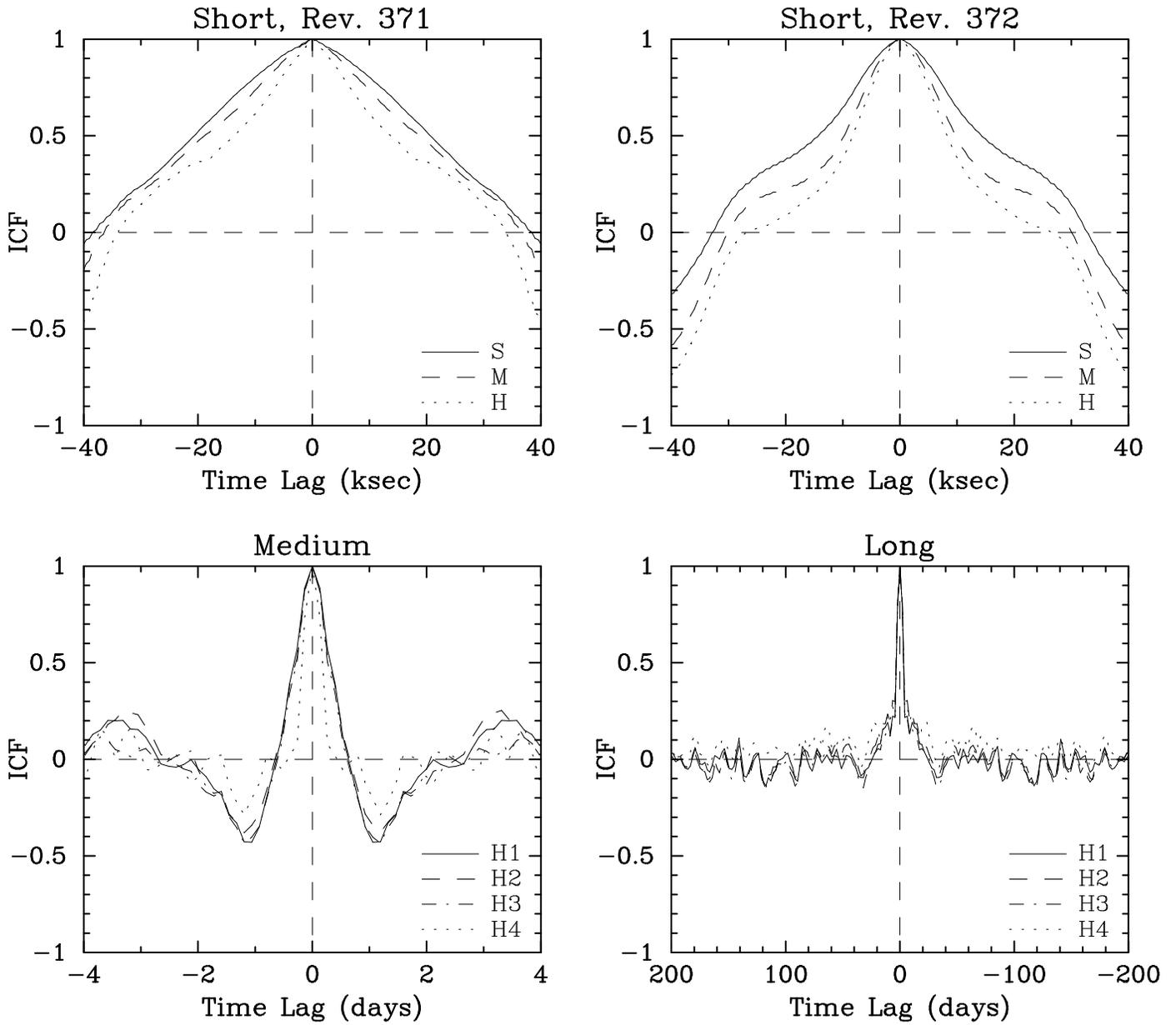}
\caption{Auto-correlation functions are shown for all time scales.}
\end{figure}

\begin{figure}    %%%%%%%%%%%%%%%%fig 6 = SR versus SUM CCF results
\epsscale{0.99}
\plotone{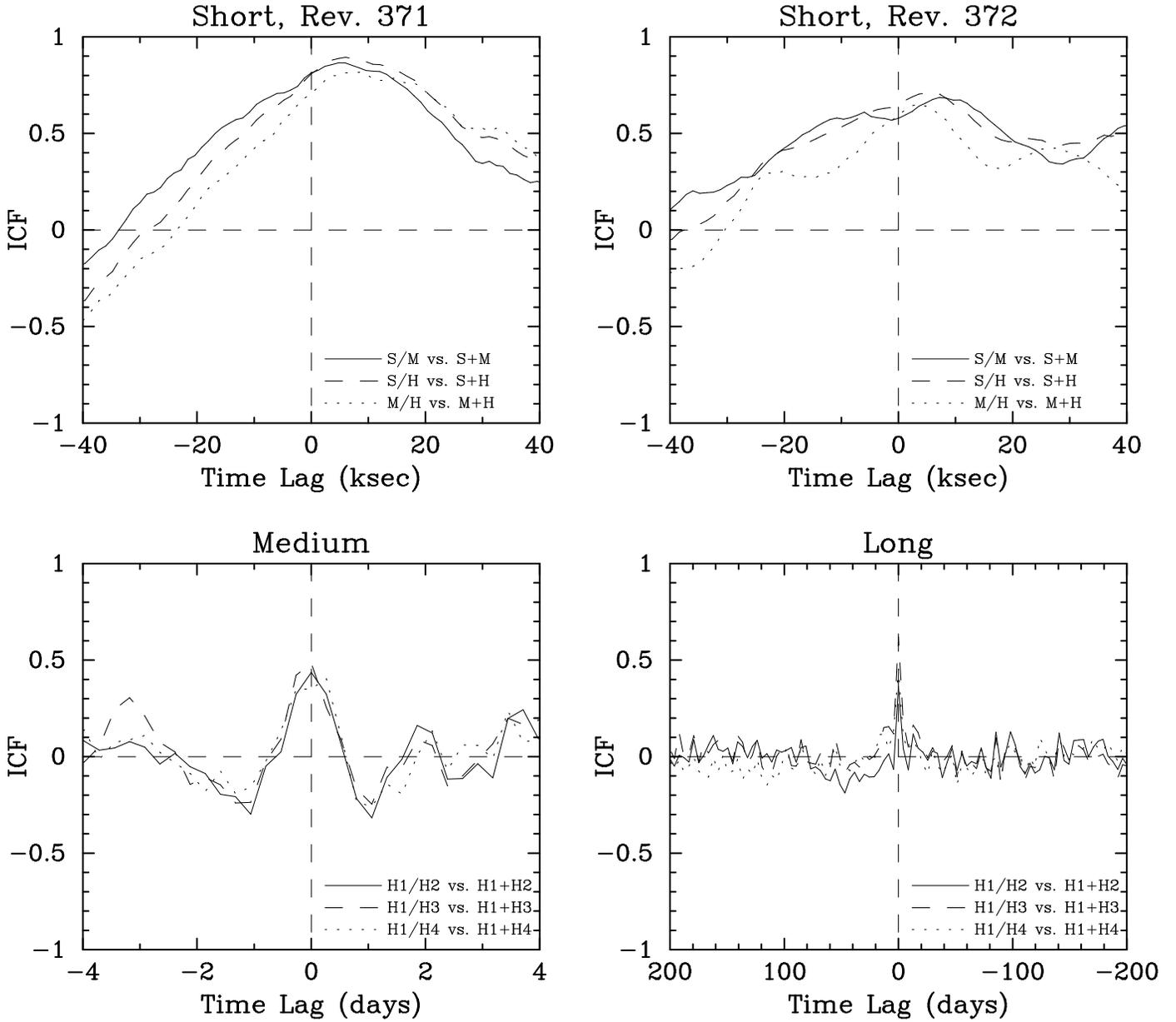}
\caption{Cross-correlation functions for the softness ratio (SR) light curves against
the summed band light curves, e.g., S/H versus (S + H). A positive 
lag indicates that variations in the summed band light curve lags those in the softness ratio light curve.}
\end{figure}

%%%%Figure for coherence  

\begin{figure}
\epsscale{0.99}
\plotone{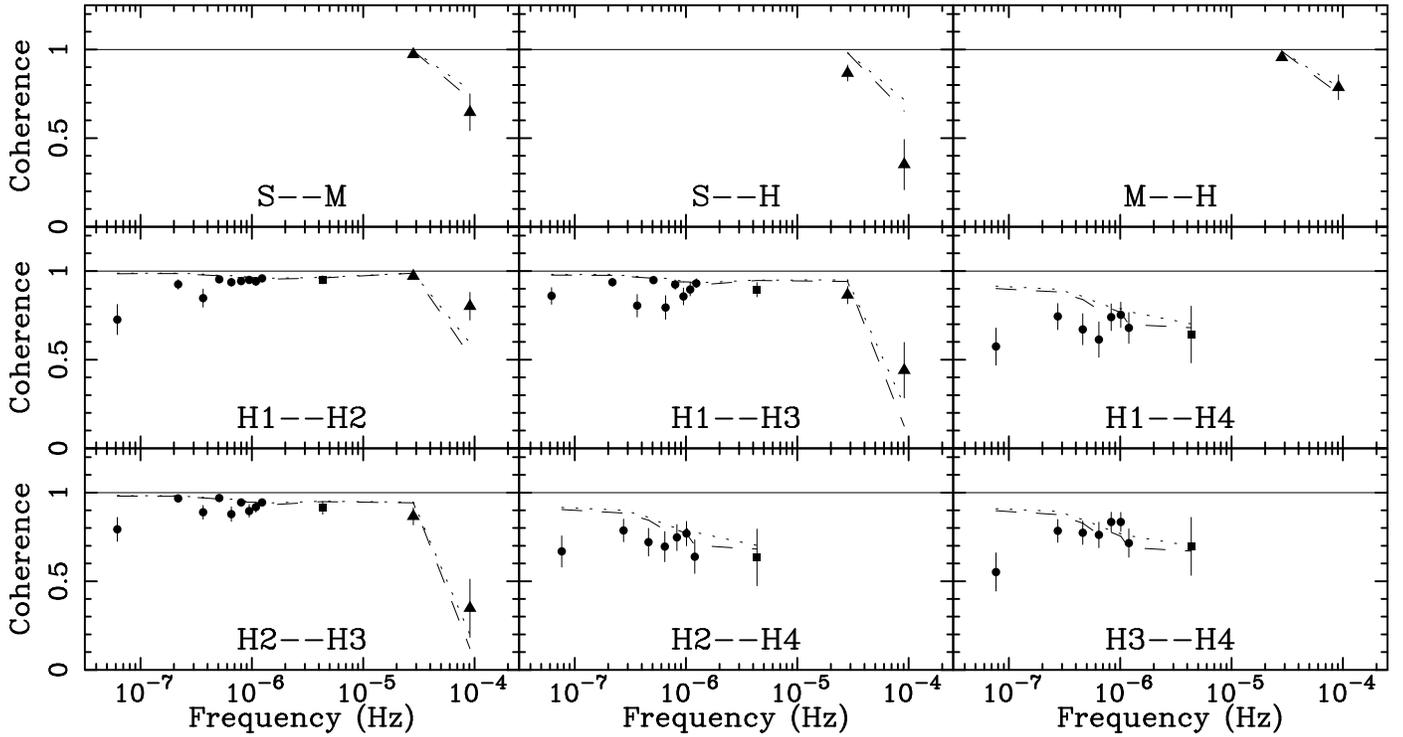}
\caption{Coherence functions for all bandpass combinations are shown.
Circles, squares, and triangles denote coherence measurements derived from the
long, medium, and short time scale data, respectively. Unity coherence is indicated by the thin
solid line. 
The dotted and dashed lines show, respectively, the 90$\%$ and 95$\%$ confidence limits  
for spurious lack of coherence due to Poisson noise. This was calculated independently for each time scale.
For most temporal frequencies, the artificial drop in
coherence is small, and especially at temporal frequencies below $\sim$10$^{-6}$ Hz, the 
deviation of the intrinsic coherence from unity is greater than the artificial drop 
at greater than 95$\%$ confidence. An exception is the highest temporal frequency point in the short-term data
(e.g., in the H2--H3 coherence plot),
where the simulations show that the drop in coherence is likely an artifact of the Poisson noise.
The intrinsic coherence has a tendency to
decrease as the energy separation of the bands increases.}
\end{figure}

\begin{figure}
\epsscale{0.59}
\plotone{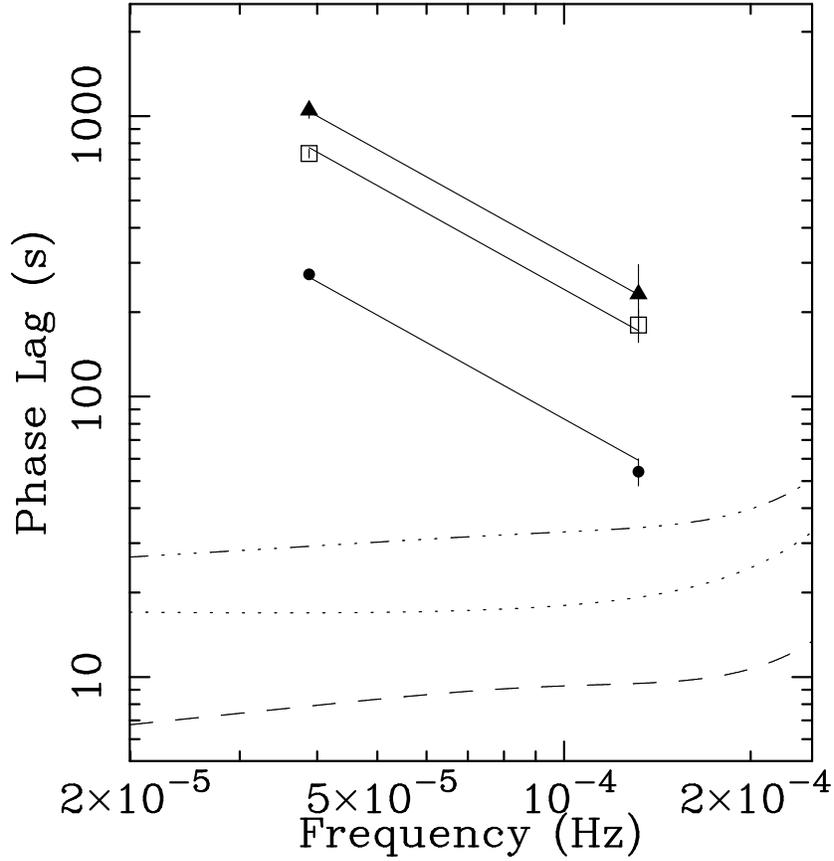}
\caption{Phase lags versus temporal frequency derived from the short term data.
Filled circles, open squares, and filled triangles show, respectively, 
the S--M, M--H, and S--H phase lags.
Phase lags are consistent with 
increasing as the energy separation of the bands increases
and as temporal frequency decreases. 
The solid lines denotes the relations 
$\phi(f)$ = 0.001 $f^{-1.2}$  (for S--M),
$\phi(f)$ = 0.003 $f^{-1.2}$  (M--H), and
$\phi(f)$ = 0.004 $f^{-1.2}$  (S--H).
The dotted, dashed, and dotted-dashed lines denote the phase lag sensitivity
limits for the S--M, M--H, and S--H bands, respectively.}
\end{figure}

\end{document}